\documentclass[12pt,english]{article}
\usepackage[T1]{fontenc}
\usepackage[latin9]{inputenc}
\usepackage{geometry}
\usepackage{refstyle}
\usepackage{amsthm}
\usepackage{amsmath}
\usepackage{amssymb}
\usepackage{esint}
\usepackage{color}
\usepackage{epsfig}
\usepackage{multirow}

\makeatletter

\AtBeginDocument{}
\RS@ifundefined{subref}
  {\def\RSsubtxt{section~}\newref{sub}{name = \RSsubtxt}}
  {}
\RS@ifundefined{thmref}
  {\def\RSthmtxt{theorem~}\newref{thm}{name = \RSthmtxt}}
  {}
\RS@ifundefined{lemref}
  {\def\RSlemtxt{lemma~}\newref{lem}{name = \RSlemtxt}}
  {}

\numberwithin{equation}{section}
\numberwithin{figure}{section}

\makeatother

\usepackage{babel}

\newcommand{\gtapprox}{\raisebox{-0.5ex}{$\,\stackrel{>}{\scriptstyle\sim}\,$}}
\newcommand{\ltapprox}{\raisebox{-0.5ex}{$\,\stackrel{<}{\scriptstyle\sim}\,$}}

\begin{document}

\begin{center}

{\huge {\bf Extracting hadron masses from fixed}}

\vspace{0.2cm}
{\huge {\bf topology simulations}}

\vspace{1.0cm}

\textbf{Arthur Dromard, Marc Wagner}

Goethe-Universit\"at Frankfurt am Main, Institut f\"ur Theoretische Physik, \\ Max-von-Laue-Stra{\ss}e 1, D-60438 Frankfurt am Main, Germany

\vspace{1.0cm}

July 16, 2014

\end{center}

\vspace{0.5cm}

\hspace{-0.6cm}\begin{tabular*}{16.6cm}{l@{\extracolsep{\fill}}r} \hline \end{tabular*}

\vspace{-0.40cm}
\begin{center} \textbf{Abstract} \end{center}
\vspace{-0.15cm}

Lattice QCD simulations tend to become stuck in a single topological sector at fine lattice spacing or when using chirally symmetric overlap quarks. In such cases physical observables differ from their full QCD counterparts by finite volume corrections. These systematic errors need to be understood on a quantitative level and possibly be removed. In this paper we extend an existing relation from the literature between two-point correlation functions at fixed and the corresponding hadron masses at unfixed topology by calculating all terms proportional to $1/V^2$ and $1/V^3$, where $V$ is the spacetime volume. Since parity is not a symmetry at fixed topology, parity mixing is comprehensively discussed. In the second part of this work we apply our equations to a simple model, quantum mechanics on a circle both for a free particle and for a square-well potential, where we demonstrate in detail, how to extract physically meaningful masses from computations or simulations at fixed topology.

\vspace{+0.15cm}
\hspace{-0.6cm}\begin{tabular*}{16.6cm}{l@{\extracolsep{\fill}}r} \hline \end{tabular*}


\newpage

\section{Introduction}

A QCD path integral includes the integration over all possible gauge or gluonic field configurations. These gauge field configurations can be classified according to their topological charge, which is integer. The numerical method to solve QCD path integrals is lattice QCD. In lattice QCD the path integral is simulated by randomly generating a representative set of gauge field configurations using Hybrid Monte Carlo (HMC) algorithms (cf.\ e.g.\ \cite{Kennedy:2006ax}). These algorithms modify a given gauge field configuration in a nearly continuous way. One of the key ideas of such a process is to generate almost exclusively gauge field configurations, which have small Euclidean action, i.e.\ which have a large weight $\propto e^{-S_\textrm{QCD,eff}}$ and, therefore, dominate the path integral (importance sampling).

To simulate a QCD path integral correctly, it is essential to sample gauge field configurations from many topological sectors. A serious problem is, however, that topological sectors are separated by large action barriers, which increase, when decreasing the lattice spacing. As a consequence, common HMC algorithms are not anymore able to frequently change the topological sector for lattice spacings $a \lesssim 0.05 \, \textrm{fm}$ \cite{Luscher:2011kk,Schaefer:2012tq}, which are nowadays still fine, but within reach.

For some lattice discretizations, e.g.\ for chirally symmetric overlap quarks, the same problem arises already at much coarser lattice spacings. Such simulations are typically performed in a single topological sector, i.e.\ at fixed topological charge \cite{Aoki:2008tq,Aoki:2012pma}, which introduces systematic errors. As an example one could mention \cite{Galletly:2006hq}, where different pion masses have been obtained for different topological charges and spacetime volumes.  Those differences have to be quantified and, if not negligible compared to statistical errors, be removed.

There are also applications, where one might fix topology on purpose, either by sorting the generated gauge field configurations with respect to their topological charge or by directly employing so-called topology fixing actions (cf.\ e.g.\ \cite{Fukaya:2005cw,Bietenholz:2005rd,Bruckmann:2009cv}). For example, when using a mixed action setup with light overlap valence and Wilson sea quarks, approximate zero modes in the valence sector are not compensated by the sea. The consequence is an ill-behaved continuum limit \cite{Cichy:2010ta,Cichy:2012vg}. Since such approximate zero modes only arise at non-vanishing topological charge, fixing topology to zero might be a way to circumvent the problem.

In view of these issues it is important to study the relation between physical quantities (i.e.\ quantities corresponding to path integrals, where gauge field configurations from many topological sectors are taken into account) and correlation functions from fixed topology simulations.

In the literature one can find an equation describing the behavior of two-point correlation functions (suited to determine hadron masses) at fixed topology, derived up to first order and in part also to second order in $1/ \chi_t V$ ($\chi_t$ is the topological susceptibility, $V$ is the spacetime volume) \cite{Brower:2003yx}, and a general discussion of higher orders for arbitrary $n$-point correlation functions at fixed topology \cite{Aoki:2007ka}. In the first more theoretically oriented part of this work (sections~\ref{SEC873} to \ref{SEC867}) we extend the calculations from \cite{Brower:2003yx} by including all terms proportional to $1/(\chi_t V)^2$ and $1/(\chi_t V)^3$. Since $\chi_t V \ltapprox 10$ in many ensembles from typical nowadays lattice QCD simulations\footnote{In particular for expensive overlap quarks as well as for very small lattice spacings, where the problem of topology freezing is most severe, one is often restricted to rather small volumes $V$, because of limited HPC resources. This in turn implies a small value of $\chi_t V$.} (cf.\ e.g.\ \cite{Aoki:2007pw,Chiu:2011dz,Cichy:2013rra,Brower:2014bqa}), fixed topology corrections of order $1/(\chi_t V)^2$ or even $1/(\chi_t V)^3$ might be sizable. Another issue we address in detail is parity mixing in fixed topology two-point correlation functions. Since parity is not a symmetry at fixed topology, masses of negative and positive parity hadrons have to be extracted from the same correlation function or matrix (in the context of the $\eta$ meson this mixing has been observed and discussed in \cite{Aoki:2007ka}). We also summarize all sources of systematic error and discuss the range of parameters (e.g.\ spatial and temporal extension of spacetime, topological charge, hadron masses), where the $1/\chi_t V$ expansions of two-point correlation functions at fixed topology are accurate approximations.

In the second part of this work (section~\ref{SEC053}) we demonstrate, how to extract hadron masses from fixed topology simulations in practice. To this end we apply the previously obtained $1/\chi_t V$ expansions of two-point correlation functions at fixed topology to a simple model, a quantum mechanical particle on a circle with and without potential. This model can be solved numerically up to arbitrary precision (there is no need to perform any simulations, only ordinary differential equations have to be solved) and, therefore, provides an ideal testbed. We have generated data points of correlation functions from many topological sectors and volumes and fit and compare different orders and versions of the previously derived correlator expansions. The results collected in various plots and tables are expected to provide helpful insights and guidelines for hadron mass determinations in quantum field theories, e.g.\ in QCD, at fixed topology (for related exploratory studies in the Schwinger model and the $O(2)$ and $O(3)$ non-linear Sigma model cf.\ \cite{Bietenholz:2011ey,Bietenholz:2012sh,Czaban:2013haa,Bautista:2014tba}).


Parts of this work have been presented at recent conferences \cite{Dromard:2013wja,Czaban:2014gva}.


\section{\label{SEC873}The partition function $Z_{Q,V}$ at fixed topology and finite spacetime volume}

In this section we calculate the dependence of the Euclidean QCD partition function at fixed topological charge $Q$ on the spacetime volume $V$, denoted as $Z_{Q,V}$, up to $\mathcal{O}(1/V^3)$.


\subsection{\label{SEC345}Calculation of the $1/V$ expansion of $Z_{Q,V}$}

The Euclidean QCD partition function at non-vanishing $\theta$ angle and finite spacetime volume $V$ is defined as
\begin{equation}
\label{EQN001} \mathcal{Z}_{\theta,V} \equiv \int DA \, D\psi \, D\bar{\psi} \, e^{-S_{E,\theta}[A,\bar{\psi},\psi]} = \sum_n e^{-E_n(\theta,V_s) T}
\end{equation}
\cite{Coleman:1978ae} with
\begin{equation}
\label{EQN002} S_{E,\theta}[A,\bar{\psi},\psi] \equiv S_E[A,\bar{\psi},\psi] + i \theta Q[A] ,
\end{equation}
where $T$ is the periodic time extension, $V_s$ the spatial volume, $V = T V_s$, $E_n(\theta,V_s)$ is the energy eigenvalue of the $n$-th eigenstate of the Hamiltonian and $S_E$ the Euclidean QCD action without $\theta$-term. Similarly, the Euclidean QCD partition function at fixed topological charge $Q$ and finite spacetime volume $V$ is defined as
\begin{equation}
\label{eq:ZQ_path} Z_{Q,V} \equiv \int DA \, D\psi \, D\bar{\psi} \, \delta_{Q,Q[A]} e^{-S_E[A,\bar{\psi},\psi]} .
\end{equation}
Using
\begin{equation}
\delta_{Q,Q[A]} = \frac{1}{2 \pi} \int_{-\pi}^{+\pi} d\theta \, e^{i (Q-Q[A]) \theta}
\end{equation}
it is easy to see that $Z_{Q,V}$ and $\mathcal{Z}_{\theta,V}$ are related by a  Fourier transform,
\begin{equation}
\label{EQN864} Z_{Q,V} = \frac{1}{2 \pi} \int_{-\pi}^{+\pi} d\theta \, e^{i Q \theta} \mathcal{Z}_{\theta,V} .
\end{equation}

One can show that $E_n(+\theta,V_s) = E_n(-\theta,V_s)$ \cite{Brower:2003yx}, which implies $(d/d\theta) E_n(\theta,V_s)|_{\theta=0} = 0$. Using this together with (\ref{EQN001}) and (\ref{EQN002}) one can express the topological susceptibility, defined as
\begin{equation}
\chi_t \equiv \lim_{V \rightarrow \infty} \frac{\langle Q^2 \rangle}{V} ,
\end{equation}
according to
\begin{equation}
\chi_t = \lim_{V_s \rightarrow \infty} \frac{E_0^{(2)}(\theta,V_s)}{V_s}\bigg|_{\theta=0} = e_0^{(2)}(\theta)\Big|_{\theta=0}
\end{equation}
(throughout this paper $X^{(n)}$ denotes the $n$-th derivative of the quantity $X$ with respect to $\theta$). Moreover, we neglect ordinary finite volume effects, i.e.\ finite volume effects not associated with fixed topology. These are expected to be suppressed exponentially with increasing spatial volume $V_s$ (cf.\ section~\ref{SEC009} for a discussion). In other words we assume $V_s$ to be sufficiently large such that $E_0(\theta,V_s) \approx e_0(\theta) V_s$, where $e_0(\theta)$ is the energy density of the vacuum.

At sufficiently large $T$ the partition function is dominated by the vacuum, i.e.\
\begin{equation}
\label{EQN003} \mathcal{Z}_{\theta,V} = e^{-E_0(\theta,V_s) T} \Big(1 + \mathcal{O}(e^{-\Delta E(\theta) T})\Big) ,
\end{equation}
where $\Delta E(\theta) = E_1(\theta,V_s) - E_0(\theta,V_s)$. The exponentially suppressed correction will be omitted in the following (cf.\ section~\ref{SEC009} for a discussion). To ease notation, we define
\begin{equation}
\label{EQN696} f(\theta) \equiv f(\theta,Q,V) \equiv e_0(\theta) - \frac{i Q \theta}{V} .
\end{equation}
Using also (\ref{EQN003}) the partition function at fixed topology (\ref{EQN864}) can be written according to
\begin{equation}
\label{EQN005} Z_{Q,V} = \frac{1}{2 \pi} \int_{-\pi}^{+\pi} d\theta \, e^{-f(\theta) V} ,
\end{equation}
where the integral (\ref{EQN005}) can be approximated by means of the saddle point method. To this end, we expand $f(\theta) V$ around its minimum $\theta_s$ and replace $\int_{-\pi}^{+\pi}$ by $\int_{-\infty}^{+\infty}$, which introduces another exponentially suppressed error (cf.\ section~\ref{SEC009} for a discussion),
\begin{equation}
\label{EQN861} Z_{Q,V} = \frac{1}{2 \pi} \int_{-\infty}^{+\infty} d\theta \, \exp\bigg(-f(\theta_s) V - \frac{f^{(2)}(\theta_s) V}{2} (\theta-\theta_s)^2 - \sum_{n=3}^\infty \frac{f^{(n)}(\theta_s) V}{n!} (\theta-\theta_s)^n\bigg) .
\end{equation}

$\theta_s$ can be determined as a power series in $1/\mathcal{E}_2 V$. Due to $E_n(+\theta,V_s) = E_n(-\theta,V_s)$, the expansion of the vacuum energy density around $\theta=0$ is
\begin{equation}
e_0(\theta) = \sum_{k=0}^\infty \frac{\mathcal{E}_{2k} \theta^{2k}}{(2k)!} \quad , \quad \mathcal{E}_k \equiv e_0^{(k)}(\theta)\Big|_{\theta=0}
\end{equation}
(note that $\mathcal{E}_2 = \chi_t$). Consequently,
\begin{equation}
\label{EQN006} f(\theta) V = \sum_{k=0}^\infty \frac{\mathcal{E}_{2k} \theta^{2k}}{(2k)!} V - i Q \theta .
\end{equation}
It is straightforward to solve the defining equation for $\theta_s$, $d/d\theta f(\theta) V|_{\theta = \theta_s} = 0$, with respect to $\theta_s$,
\begin{equation}
\label{eq:theta_s} \theta_{s} = i \bigg(\frac{1}{\mathcal{E}_2 V} Q + \frac{1}{(\mathcal{E}_2 V)^3} \frac{\mathcal{E}_4}{6 \mathcal{E}_2} Q^3\bigg)
+ \mathcal{O}\bigg(\frac{1}{(\mathcal{E}_2 V)^5}\bigg) \footnote{Throughout this work errors in $1 / \mathcal{E}_2 V$ are proportional to either $1 / (\mathcal{E}_2 V)^4$ or $1 / (\mathcal{E}_2 V)^5$. For errors proportional to $1 / (\mathcal{E}_2 V)^4$ we also keep track of powers of $Q$, e.g.\ we distinguish $\mathcal{O}(1 / (\mathcal{E}_2 V)^4)$ and $\mathcal{O}(Q^2 / (\mathcal{E}_2 V)^4)$, etc. For errors proportional to $1 / (\mathcal{E}_2 V)^5$, we do not show powers of $Q$, i.e.\ we just write $\mathcal{O}(1 / (\mathcal{E}_2 V)^5)$. We also estimate $\mathcal{E}_n / \mathcal{E}_2 = \mathcal{O}(1)$, for which numerical support can be found in \cite{Bonati:2013tt}} .
\end{equation}

Finally the saddle point method requires to deform the contour of integration to pass through the saddle point, which is just a constant shift of the real axis by the purely imaginary $\theta_s$. We introduce the real coordinate $s \equiv (\theta - \theta_s) (f^{(2)}(\theta_s) V)^{1/2}$ parameterizing the shifted contour of integration yielding
\begin{equation}
\label{EQN936} Z_{Q,V} = \frac{e^{-f(\theta_s) V}}{2 \pi (f^{(2)}(\theta_s) V)^{1/2}} \int_{-\infty}^{+\infty} ds \, \exp\bigg(-\frac{1}{2} s^2 - \sum_{n=3}^\infty \frac{f^{(n)}(\theta_s) V}{n! (f^{(2)}(\theta_s) V)^{n/2}} s^n\bigg) .
\end{equation}
After defining
\begin{equation}
||h(s)|| \equiv \frac{1}{\sqrt{2 \pi}} \int_{-\infty}^{+\infty} ds \, e^{-s^2/2} h(s)
\end{equation}
a more compact notation for the result (\ref{EQN936}) is
\begin{equation}
\label{eq: Saddle point approximation} Z_{Q,V} = \frac{e^{-f(\theta_s) V}}{(2 \pi f^{(2)}(\theta_s) V)^{1/2}} \underbrace{\bigg\Vert \exp\bigg(-\sum_{n=3}^\infty \frac{f^{(n)}(\theta_s) V}{n! (f^{(2)}(\theta_s) V)^{n/2}} s^n\bigg)\bigg\Vert}_{\equiv G} ,
\end{equation}
where $G$ can also be written as
\begin{equation}
\label{EQN007} G = 1 + \sum_{k=1}^\infty \frac{(-1)^k}{k!} \bigg\Vert\bigg(\sum_{n=3}^\infty \frac{f^{(n)}(\theta_{s}) V}{n! (f^{(2)}(\theta_s) V)^{n/2}} s^n\bigg)^k\bigg\Vert .
\end{equation}

We now insert $f(\theta) V$ and $\theta_s$ (eqs.\ (\ref{EQN006}) and (\ref{eq:theta_s})) and perform the integration over $s$ order by order in $1 / \mathcal{E}_2 V$ (note that $\theta_s \sim 1/\mathcal{E}_2 V$). To this end we use the relations
\begin{equation}
\label{EQN008} \begin{aligned}
 & f^{(2n)}(\theta_s) V = \sum_{l=n}^\infty \frac{\mathcal{E}_{2l} V}{(2l-2n)!} \theta_s^{2l-2n} \quad , \quad n = 1,2,\ldots \\
 & f^{(2n-1)}(\theta_s) V = \sum_{l=n}^\infty \frac{\mathcal{E}_{2l} V}{(2l-2n+1)!} \theta_s^{2l-2n+1} \quad , \quad n = 2,3,\ldots \\
 & ||s^{2n-1}|| = 0 \quad , \quad ||s^{2n}|| = (2n-1)!! = \frac{(2n)!}{2^n n!} = 1 \times 3 \times 5 \times \ldots \times (2n-1) \quad , \quad n = 0,1,\ldots
\end{aligned}
\end{equation}
The terms in (\ref{EQN007}) are
\begin{itemize}
\item for $k = 1$ proportional to $1 / (\mathcal{E}_2 V)^{n/2-1}$,

\item for $k = 2$ proportional to $1 / (\mathcal{E}_2 V)^{(n_1+n_2)/2-2}$,

\item for $k = 3$ proportional to $1 / (\mathcal{E}_2 V)^{(n_1+n_2+n_3)/2-3}$, ...
\end{itemize}
Moreover, $n$, $n_1+n_2$, $n_1+n_2+n_3$, ... have to be even, otherwise the corresponding term in (\ref{EQN007}) vanishes, due to (\ref{EQN008}). Finally every odd $n$ and $n_j$ contributes in leading order in $\theta_s$ one power of $\theta_s \sim 1/\mathcal{E}_2 V$. Therefore, up to $\mathcal{O}(1/(\mathcal{E}_2 V)^3)$ it is sufficient to consider the following terms:
\begin{itemize}
\item $k=1$, $n=4$:
\begin{equation}
\bigg\Vert \frac{f^{(4)}(\theta_s) V}{4! (f^{(2)}(\theta_s) V)^2} s^4 \bigg\Vert =
\frac{1}{\mathcal{E}_2 V} \frac{\mathcal{E}_4}{8 \mathcal{E}_2} +
\frac{1}{\mathcal{E}_2 V} \bigg(\frac{\mathcal{E}_6}{16 \mathcal{E}_2} - \frac{\mathcal{E}_4^2}{8 \mathcal{E}_2^2}\bigg) \theta_{s}^2 +
\mathcal{O}\bigg(\frac{1}{(\mathcal{E}_2 V)^5}\bigg) .
\end{equation}

\item $k=1$, $n=6$:
\begin{equation}
\bigg\Vert \frac{f^{(6)}(\theta_s) V}{6! (f^{(2)}(\theta_s) V)^3} s^6 \bigg\Vert =
\frac{1}{(\mathcal{E}_2 V)^2} \frac{\mathcal{E}_6}{48 \mathcal{E}_2} +
\mathcal{O} \bigg(\frac{1}{(\mathcal{E}_2 V)^2} \theta_s^2\bigg) .
\end{equation}

\item $k=1$, $n=8$:
\begin{equation}
\bigg\Vert \frac{f^{(8)}(\theta_s) V}{8! (f^{(2)}(\theta_s) V)^4} s^8 \bigg\Vert =
\frac{1}{(\mathcal{E}_2 V)^3} \frac{\mathcal{E}_8}{384 \mathcal{E}_2} +
\mathcal{O}\bigg(\frac{1}{(\mathcal{E}_2 V)^5}\bigg) .
\end{equation}

\item $k=1$, $n=10$:
\begin{equation}
\bigg\Vert \frac{f^{(10)}(\theta_s) V}{10! (f^{(2)}(\theta_s) V)^5} s^{10}\bigg\Vert =
\mathcal{O}\bigg(\frac{1}{(\mathcal{E}_2 V)^4}\bigg) .
\end{equation}


\item $k=2$, $n_1=n_2=3$:
\begin{equation}
\bigg\Vert \frac{(f^{(3)}(\theta_s) V)^2}{(3!)^2 (f^{(2)}(\theta_s) V)^3} s^6 \bigg\Vert =
\frac{1}{\mathcal{E}_2 V} \frac{5 \mathcal{E}_4^2}{12 \mathcal{E}_2^2} \theta_s^2 +
\mathcal{O}\bigg(\frac{1}{(\mathcal{E}_2 V)^5}\bigg) .
\end{equation}

\item $k=2$, $n_1=3$, $n_2=5$:
\begin{equation}
2 \times \bigg\Vert \frac{(f^{(3)}(\theta_s) V) (f^{(5)}(\theta_s) V)}{3! 5! (f^{(2)}(\theta_s) V)^4} s^8 \bigg\Vert =
\mathcal{O}\bigg(\frac{1}{(\mathcal{E}_2 V)^2} \theta_s^2\bigg) .
\end{equation}

\item $k=2$, $n_1=n_2=4$:
\begin{equation}
\bigg\Vert \frac{(f^{(4)}(\theta_s) V)^2}{(4!)^2 (f^{(2)}(\theta_s) V)^4} s^8\bigg\Vert =
\frac{1}{(\mathcal{E}_2 V)^2} \frac{35 \mathcal{E}_4^2}{192 \mathcal{E}_2^2} +
\mathcal{O}\bigg(\frac{1}{(\mathcal{E}_2 V)^2} \theta_s^2\bigg) .
\end{equation}

\item $k=2$, $n_1=4$, $n_2=6$:
\begin{equation}
2 \times \bigg\Vert \frac{(f^{(4)}(\theta_s) V) (f^{(6)}(\theta_s) V)}{4! 6! (f^{(2)}(\theta_s) V)^5} s^{10} \bigg\Vert =
\frac{1}{(\mathcal{E}_2 V)^3} \frac{7 \mathcal{E}_4 \mathcal{E}_6}{64 \mathcal{E}_2^2} +
\mathcal{O}\bigg(\frac{1}{(\mathcal{E}_2 V)^5}\bigg) .
\end{equation}

\item $k=2$, $n_1=4$, $n_2=8$:
\begin{equation}
2 \times \bigg\Vert \frac{(f^{(4)}(\theta_s) V) (f^{(8)}(\theta_s) V)}{4! 8! (f^{(2)}(\theta_s) V)^6} s^{12} \bigg\Vert =
\mathcal{O}\bigg(\frac{1}{(\mathcal{E}_2 V)^4}\bigg) .
\end{equation}

\item $k=2$, $n_1=n_2=6$:
\begin{equation}
\bigg\Vert \frac{(f^{(6)}(\theta_s) V)^2}{(6!)^2 (f^{(2)}(\theta_s) V)^6} s^{12}\bigg\Vert =
\mathcal{O}\bigg(\frac{1}{(\mathcal{E}_2 V)^4}\bigg) .
\end{equation}


\item $k=3$, $n_1=n_2=3$, $n_3=4$:
\begin{equation}
3 \times \bigg\Vert \frac{(f^{(3)}(\theta_s) V)^2 (f^{(4)}(\theta_s) V)}{(3!)^2 4! (f^{(2)}(\theta_s) V)^5} s^{10} \bigg\Vert =
\mathcal{O}\bigg(\frac{1}{(\mathcal{E}_2 V)^2} \theta_s^2\bigg) .
\end{equation}

\item $k=3$, $n_1=n_2=n_3=4$:
\begin{equation}
\bigg\Vert \frac{(f^{(4)}(\theta_s) V)^3}{(4!)^3 (f^{(2)}(\theta_s) V)^6} s^{12} \bigg\Vert =
\frac{1}{(\mathcal{E}_2 V)^3} \frac{385 \mathcal{E}_4^3}{512 \mathcal{E}_2^3} +
\mathcal{O}\bigg(\frac{1}{(\mathcal{E}_2 V)^5}\bigg) .
\end{equation}

\item $k=3$, $n_1=n_2=4$, $n_3=6$:
\begin{equation}
3 \times \bigg\Vert \frac{(f^{(4)}(\theta_s) V)^2 (f^{(6)}(\theta_s) V)}{(4!)^2 6! (f^{(2)}(\theta_s) V)^7} s^{14} \bigg\Vert =
\mathcal{O}\bigg(\frac{1}{(\mathcal{E}_2 V)^4}\bigg) .
\end{equation}


\item $k=4$, $n_1=n_2=n_3=n_4=4$:
\begin{equation}
\bigg\Vert \frac{(f^{(4)}(\theta_s) V)^4}{(4!)^4 (f^{(2)}(\theta_s) V)^8} s^{16} \bigg\Vert =
\mathcal{O}\bigg(\frac{1}{(\mathcal{E}_2 V)^4}\bigg) .
\end{equation}
\end{itemize}
Inserting these expressions into (\ref{EQN007}) leads to
\begin{equation}
\begin{aligned}
 & G = 1 + \frac{1}{\mathcal{E}_2 V} \bigg(-\frac{\mathcal{E}_4}{8 \mathcal{E}_2} + \bigg(-\frac{\mathcal{E}_6}{16 \mathcal{E}_2} + \frac{\mathcal{E}_4^{2}}{3 \mathcal{E}_2^2}\bigg) \theta_s^2\bigg) + \frac{1}{(\mathcal{E}_2 V)^2} \bigg(-\frac{\mathcal{E}_6}{48 \mathcal{E}_2} + \frac{35 \mathcal{E}_4^2}{384 \mathcal{E}_2^2}\bigg) \\
 & \hspace{1.4cm} + \frac{1}{(\mathcal{E}_2 V)^3} \bigg(-\frac{\mathcal{E}_8}{384 \mathcal{E}_2} + \frac{7 \mathcal{E}_4 \mathcal{E}_6}{128 \mathcal{E}_2^2} - \frac{385 \mathcal{E}_{4}^3}{3072 \mathcal{E}_2^3}\bigg) + \mathcal{O}\bigg(\frac{1}{(\mathcal{E}_2 V)^4} \ , \ \frac{1}{(\mathcal{E}_2 V)^2} \theta_s^2\bigg)
\end{aligned}
\end{equation}
and, after inserting the expansion of $\theta_s$ (\ref{eq:theta_s}), yields
\begin{equation}
\label{EQN704} \begin{aligned}
 & G = 1 - \frac{1}{\mathcal{E}_2 V} \frac{\mathcal{E}_4}{8 \mathcal{E}_2} + \frac{1}{(\mathcal{E}_2 V)^2} \bigg(-\frac{\mathcal{E}_6}{48 \mathcal{E}_2} + \frac{35 \mathcal{E}_4^2}{384 \mathcal{E}_2^2}\bigg) \\
 & \hspace{1.4cm} + \frac{1}{(\mathcal{E}_2 V)^3} \bigg(-\frac{\mathcal{E}_8}{384 \mathcal{E}_2} + \frac{7 \mathcal{E}_4 \mathcal{E}_6}{128 \mathcal{E}_2^2} - \frac{385 \mathcal{E}_4^3}{3072 \mathcal{E}_2^3} + \bigg(\frac{\mathcal{E}_6}{16 \mathcal{E}_2} - \frac{\mathcal{E}_4^2}{3 \mathcal{E}_2^2}\bigg) Q^2\bigg) \\
 & \hspace{1.4cm} + \mathcal{O}\bigg(\frac{1}{(\mathcal{E}_2 V)^4} \ , \ \frac{1}{(\mathcal{E}_2 V)^4} Q^2\bigg) .
\end{aligned}
\end{equation}

The remaining terms in (\ref{eq: Saddle point approximation}) expressed in powers of $V$ are
\begin{equation}
\label{EQN705} f(\theta_s) V = \mathcal{E}_0 V + \frac{1}{\mathcal{E}_2 V} \frac{1}{2} Q^2 + \frac{1}{(\mathcal{E}_2 V)^3} \frac{\mathcal{E}_{4}}{24 \mathcal{E}_2} Q^4 + \mathcal{O}\bigg(\frac{1}{(\mathcal{E}_2 V)^5}\bigg)
\end{equation}
and
\begin{equation}
\label{EQN706} f^{(2)}(\theta_s) V = \mathcal{E}_2 V \bigg(1 - \frac{1}{(\mathcal{E}_2 V)^2} \frac{\mathcal{E}_4}{2 \mathcal{E}_2} Q^2 + \mathcal{O}\bigg(\frac{1}{(\mathcal{E}_2 V)^4} Q^4\bigg)\bigg) .
\end{equation}
Combining (\ref{eq: Saddle point approximation}), (\ref{EQN704}), (\ref{EQN705}) and (\ref{EQN706}) yields the final result for $Z_{Q,V}$,
\begin{equation}
\label{EQN010} \begin{aligned}
 & Z_{Q,V} = \frac{1}{\sqrt{2 \pi \mathcal{E}_2 V}} \bigg(\exp\bigg(-E_0(0,V_s) T - \frac{1}{\mathcal{E}_2 V} \frac{1}{2} Q^2 - \frac{1}{(\mathcal{E}_2 V)^3} \frac{\mathcal{E}_4}{24 \mathcal{E}_2} Q^4\bigg) \\
 & \hspace{2.1cm} \bigg(1 - \frac{1}{(\mathcal{E}_2 V)^2} \frac{\mathcal{E}_4}{2 \mathcal{E}_2} Q^2\bigg)^{-1/2} G \\
 & \hspace{1.4cm} + \mathcal{O}\bigg(\frac{1}{(\mathcal{E}_2 V)^4} Q^4\bigg)\bigg) = \\
 & \hspace{0.7cm} = \frac{1}{\sqrt{2 \pi \mathcal{E}_2 V}} \bigg(\exp\bigg(-E_0(0,V_s) T - \frac{1}{\mathcal{E}_2 V} \frac{1}{2} Q^2 - \frac{1}{(\mathcal{E}_2 V)^3} \frac{\mathcal{E}_4}{24 \mathcal{E}_2} Q^4\bigg) \\
 & \hspace{2.1cm} \bigg(1 - \frac{1}{(\mathcal{E}_2 V)^2} \frac{\mathcal{E}_4}{2 \mathcal{E}_2} Q^2\bigg)^{-1/2} \\
 & \hspace{2.1cm} \bigg(1 - \frac{1}{\mathcal{E}_2 V} \frac{\mathcal{E}_4}{8 \mathcal{E}_2} + \frac{1}{(\mathcal{E}_2 V)^2} \bigg(-\frac{\mathcal{E}_6}{48 \mathcal{E}_2} + \frac{35 \mathcal{E}_4^2}{384 \mathcal{E}_2^2}\bigg) \\
 & \hspace{2.8cm} + \frac{1}{(\mathcal{E}_2 V)^3} \bigg(-\frac{\mathcal{E}_8}{384 \mathcal{E}_2} + \frac{7 \mathcal{E}_4 \mathcal{E}_6}{128 \mathcal{E}_2^2} - \frac{385 \mathcal{E}_4^3}{3072 \mathcal{E}_2^3} + \bigg(\frac{\mathcal{E}_6}{16 \mathcal{E}_2} - \frac{\mathcal{E}_4^2}{3 \mathcal{E}_2^2}\bigg) Q^2\bigg) \\
 & \hspace{1.4cm} + \mathcal{O}\bigg(\frac{1}{\mathcal{E}_2^4 V^4} \ , \ \frac{1}{\mathcal{E}_2^4 V^4} Q^2 \ , \ \frac{1}{\mathcal{E}_2^4 V^4} Q^4\bigg)\bigg) .
\end{aligned}
\end{equation}


\subsection{Comparison with \cite{Brower:2003yx}}

It is easy to see that equation (2.16) derived in \cite{Brower:2003yx},
\begin{equation}
Z_Q = \frac{1}{\sqrt{2 \pi \beta V \chi_t}} \exp\bigg(-\frac{Q^2}{2 \beta V \chi_t}\bigg) \bigg(1 + \mathcal{O}\bigg(\frac{\gamma}{\beta V}\bigg)\bigg) ,
\end{equation}
is contained in our result (\ref{EQN010}), after changing notation according to $\beta V \rightarrow V$ and $\chi_t \rightarrow \mathcal{E}_2$ (in \cite{Brower:2003yx} $\mathcal{E}_0 = 0$ has been assumed and $\gamma \propto \mathcal{E}_4$ is a constant).


\section{\label{SEC459}Two-point correlation functions $C_{Q,V}(t)$ at fixed topology and finite spacetime volume}

In this section we derive a relation between physical hadron masses (i.e.\ at unfixed topology and $\theta = 0$) and the corresponding two-point correlation functions at fixed topological charge $Q$ and finite spacetime volume $V$, denoted as $C_{Q,V}(t)$, up to $\mathcal{O}(1/V^3)$.


\subsection{\label{SEC346}Calculation of the $1/V$ expansion of $C_{Q,V}(t)$}

Two-point correlation functions at fixed topological charge $Q$ and finite spacetime volume $V$ are defined as
\begin{equation}
C_{Q,V}(t) \equiv \frac{1}{Z_{Q,V}} \int DA \, D\psi \, D\bar{\psi} \, \delta_{Q,Q[A]} O^\dagger(t) O(0) e^{-S_E[A,\bar{\psi},\psi]} .
\end{equation}
$O$ denotes a suitable hadron creation operator, for example for the charged pion $\pi^+$ a common choice is
\begin{equation}
\label{EQN237} O \equiv \frac{1}{\sqrt{V_s}} \int d^3r \, \bar{d}(\mathbf{r}) \gamma_5 u(\mathbf{r})
\end{equation}
(cf.\ e.g.\ \cite{Weber:2013eba} for an introduction in lattice hadron spectroscopy and the construction of hadron creation operators). $C_{Q,V}(t)$ is related to a corresponding two-point correlation function at non-vanishing $\theta$ angle and finite spacetime volume $V$ defined as
\begin{equation}
\mathcal{C}_{\theta,V}(t) \equiv \frac{1}{\mathcal{Z}_{\theta,V}} \int DA \, D\psi \, D\bar{\psi} \, O^\dagger(t) O(0) e^{-S_{E,\theta}[A,\bar{\psi},\psi]}
\end{equation}
via a Fourier transform,
\begin{equation}
\label{EQN415} C_{Q,V}(t) = \frac{1}{2 \pi Z_{Q,V}} \int_{-\pi}^{+\pi} d\theta \, \mathcal{Z}_{\theta,V} \mathcal{C}_{\theta,V}(t) e^{i Q \theta} .
\end{equation}

$\mathcal{C}_{\theta,V}(t)$ can be expressed in terms of energy eigenstates $|n ; \theta , V_s \rangle$ and eigenvalues,
\begin{equation}
\label{EQN597} \mathcal{C}_{\theta,V}(t) \mathcal{Z}_{\theta,V} = \sum_{n,m} \Big|\langle m ; \theta , V_s | O |n ; \theta , V_s \rangle\Big|^2 e^{-E_m(\theta,V_s) t} e^{-E_n(\theta,V_s)(T-t)} .
\end{equation}
When applied to the vacuum $| 0 ; \theta , V_s \rangle$, the hadron creation operator $O$ creates a state, which has the quantum numbers of the hadron of interest $H$, which are assumed to be not identical to those of the vacuum, even at $\theta \neq 0$. These states are denoted by $| H , n ; \theta , V_s \rangle$, the corresponding eigenvalues by $E_{H,n}(\theta,V_s)$. $H$ is typically the lowest state in that sector\footnote{Note that parity is not a symmetry at $\theta \neq 0$. Therefore, states with defined parity at $\theta = 0$, which have lighter parity partners (e.g.\ positive parity mesons), have to be treated and extracted as excited states at $\theta \neq 0$ and, consequently, also at fixed topology. This more complicated case is discussed in section~\ref{SEC692}.}, i.e.\ $| H , 0 ; \theta , V_s \rangle$ with mass $M_H(\theta) \equiv E_{H,0}(\theta,V_s) - E_0(\theta,V_s)$ (in this section we again neglect ordinary finite volume effects, i.e.\ finite volume effects not associated with fixed topology; cf.\ section~\ref{SEC009} for a discussion). Using this notation one can rewrite (\ref{EQN597}) according to
\begin{equation}
\label{EQN856} \begin{aligned}
 & \mathcal{C}_{\theta,V}(t) \mathcal{Z}_{\theta,V} = \\
 & \hspace{0.7cm} = \alpha(\theta,V_s) e^{-E_0(\theta,V_s) T} e^{-M_H(\theta) t} + \mathcal{O}(e^{-E_0(\theta,V_s) T} e^{-M_H^\ast(\theta) t}) + \mathcal{O}(e^{-E_0(\theta,V_s) T} e^{-M_H(\theta) (T-t)}) = \\
 & \hspace{0.7cm} = \alpha(\theta,V_s) e^{-E_0(\theta,V_s) T} e^{-M_H(\theta) t} \Big(1 + \mathcal{O}(e^{-(M_H^\ast(\theta) - M_H(\theta)) t}) + \mathcal{O}(e^{-M_H(\theta) (T-2 t)})\Big) ,
\end{aligned}
\end{equation}
where $\alpha(\theta,V_s) \equiv |\langle H, 0 ; \theta , V_s | O | 0 ; \theta , V_s \rangle|^2$ and $M_H^\ast(\theta) \equiv E_{H,1}(\theta,V_s)-E_0(\theta,V_s)$ is the mass of the first excitation with the quantum numbers of $H$.

For suitably normalized hadron creation operators $O$, e.g.\ operators
\begin{equation}
O \equiv \frac{1}{\sqrt{V_s}} \int d^3r \, O'(\mathbf{r}) ,
\end{equation}
where $O'(\mathrm{r})$ is a local operator, i.e.\ an operator exciting quark and gluon fields only at or close to $\mathbf{r}$, $\alpha$ is independent of $V_s$, i.e.\ $\alpha = \alpha(\theta)$. Moreover, for operators $O$ respecting either $P O P = +O$ or $P O P = -O$, i.e.\ operators with defined parity $P$, one can show $\alpha(+\theta) = \alpha(-\theta)$ by using $P | n ; -\theta , V_s \rangle = \eta_n(\theta , V_s) | n ; +\theta , V_s \rangle$, where $\eta_n(\theta , V_s)$ is a non-unique phase. In the following we assume that $O$ is suitably normalized and has defined parity. Then $\alpha(\theta)$ can be written as a power series around $\theta = 0$ according to
\begin{equation}
\label{EQN999} \alpha(\theta) = \sum_{k=0}^\infty \frac{\alpha^{(2k)}(0) \theta^{2k}}{(2k)!} = \alpha(0) \exp\bigg(\underbrace{\ln\bigg(\sum_{k=0}^\infty \frac{\alpha^{(2k)}(0) \theta^{2k}}{(2k)! \alpha(0)}\bigg)}_{\equiv -\beta(\theta) = -\sum_{k=1}^\infty \frac{\beta^{(2k)}(0) \theta^{2k}}{(2k)!}}\bigg) .
\end{equation}
Inserting $\alpha(\theta)$ in (\ref{EQN856}) and neglecting exponentially suppressed corrections (cf.\ section~\ref{SEC009} for a discussion) leads to
\begin{equation}
\mathcal{C}_{\theta,V}(t) \mathcal{Z}_{\theta,V} = \alpha(0) e^{-(e_0(\theta) V + M_H(\theta) t + \beta(\theta))} .
\end{equation}

In analogy to (\ref{EQN696}) we define
\begin{equation}
f_{\mathcal{C}}(\theta) \equiv f_{\mathcal{C}}(\theta,Q,V) \equiv e_0(\theta) + \frac{M_H(\theta) t + \beta(\theta) - i Q \theta}{V} .
\end{equation}
For two-point correlation functions at fixed topology we then arrive at a similar form as for $Z_{Q,V}$ (eq.\ (\ref{EQN005})),
\begin{equation}
C_{Q,V}(t) Z_{Q,V} = \frac{\alpha(0)}{2 \pi} \int_{-\pi}^{+\pi} d\theta \, e^{-f_{\mathcal{C}}(\theta) V} .
\end{equation}
With
\begin{equation}
\label{EQN995}\mathcal{F}_{2k} \equiv \mathcal{E}_{2k} + \frac{M_H^{(2k)}(0) t + \beta^{(2k)}(0)}{V} = \mathcal{E}_{2k} \bigg(1 + \frac{x_{2k}}{\mathcal{E}_{2k} V}\bigg) \quad , \quad x_{2k} \equiv M_H^{(2k)}(0) t + \beta^{(2k)}(0)
\end{equation}
the expansion of the exponent is
\begin{equation}
\label{EQN079} f_{\mathcal{C}}(\theta) V = \sum_{k=0}^\infty \frac{\mathcal{F}_{2k} \theta^{2k}}{(2k)!} V - i Q \theta .
\end{equation}
Up to $\mathcal{O}(1/(\mathcal{E}_2 V)^4)$ its minimum can easily be obtained by using (\ref{eq:theta_s}),
\begin{equation}
\label{EQN357} \begin{aligned}
 & \theta_{s,\mathcal{C}} = i \bigg(\frac{1}{\mathcal{F}_2 V} Q + \frac{1}{(\mathcal{F}_2 V)^3} \frac{\mathcal{F}_4}{6 \mathcal{F}_2} Q^3\bigg) + \mathcal{O}\bigg(\frac{1}{(\mathcal{F}_2 V)^5}\bigg) \\
 & \hspace{0.7cm} = i \bigg(\frac{1}{\mathcal{E}_2 V (1 + x_2 / \mathcal{E}_2 V)} Q + \frac{1}{(\mathcal{E}_2 V)^3} \frac{\mathcal{E}_4 (1 + x_4 / \mathcal{E}_4 V)}{6 \mathcal{E}_2 (1 + x_2 / \mathcal{E}_2 V)^4} Q^3\bigg) + \mathcal{O}\bigg(\frac{1}{(\mathcal{E}_2 V)^5}\bigg) .
\end{aligned}
\end{equation}
$C_{Q,V}(t) Z_{Q,V}$ can be written in the same form as $Z_{Q,V}$ (eq.\ (\ref{eq: Saddle point approximation})),
\begin{equation}
\label{eq:saddle point C_Q} C_{Q,V}(t) Z_{Q,V} = \frac{\alpha(0) e^{-f_{\mathcal{C}}(\theta_{s,\mathcal{C}}) V}}{\sqrt{2 \pi} (f_{\mathcal{C}}^{(2)}(\theta_{s,\mathcal{C}}) V)^{1/2}} \underbrace{\bigg\Vert \exp\bigg(-\sum_{n=3}^\infty \frac{f_{\mathcal{C}}^{(n)}(\theta_{s,\mathcal{C}}) V}{n! (f_{\mathcal{C}}^{(2)}(\theta_{s,\mathcal{C}}) V)^{n/2}} s^n\bigg)\bigg\Vert}_{\equiv G_\mathcal{C}} .
\end{equation}
Using (\ref{EQN704}) and (\ref{EQN010}) yields an explicit expression up to $\mathcal{O}(1 / (\mathcal{E}_2 V)^3)$,
\begin{equation}
\label{EQN624} \begin{aligned}
 & C_{Q,V}(t) Z_{Q,V} = \frac{\alpha(0)}{\sqrt{2 \pi \mathcal{F}_2 V}} \bigg(\exp\bigg(-\mathcal{F}_0 V - \frac{1}{\mathcal{F}_2 V} \frac{1}{2} Q^2 - \frac{1}{(\mathcal{F}_2 V)^3} \frac{\mathcal{F}_{4}}{24 \mathcal{F}_2} Q^4\bigg) \\
 & \hspace{2.1cm} \bigg(1 - \frac{1}{(\mathcal{F}_2 V)^2} \frac{\mathcal{F}_4}{2 \mathcal{F}_2} Q^2\bigg)^{-1/2} G_\mathcal{C} + \mathcal{O}\bigg(\frac{1}{(\mathcal{E}_2 V)^4} Q^4\bigg)\bigg) = \\
 & \hspace{0.7cm} = \frac{1}{\sqrt{2 \pi \mathcal{E}_2 V}} \frac{\alpha(0)}{\sqrt{1 + x_2 / \mathcal{E}_2 V}} \\
 & \hspace{1.4cm} \bigg(\exp\bigg(-E_0 T - M_H(0) t - \frac{1}{\mathcal{E}_2 V (1 + x_2 / \mathcal{E}_2 V)} \frac{1}{2} Q^2 \\
 & \hspace{2.8cm} - \frac{1}{(\mathcal{E}_2 V)^3} \frac{\mathcal{E}_4 (1 + x_4 / \mathcal{E}_4 V)}{24 \mathcal{E}_2 (1 + x_2 / \mathcal{E}_2 V)^4} Q^4\bigg) \\
 & \hspace{2.1cm} \bigg(1 - \frac{1}{(\mathcal{E}_2 V)^2} \frac{\mathcal{E}_4 (1 + x_4 / \mathcal{E}_4 V)}{2 \mathcal{E}_2 (1 + x_2 / \mathcal{E}_2 V)^3} Q^2\bigg)^{-1/2} G_\mathcal{C} + \mathcal{O}\bigg(\frac{1}{(\mathcal{E}_2 V)^4} Q^4\bigg)\bigg)
\end{aligned}
\end{equation}
with
\begin{equation}
\label{EQN387} \begin{aligned}
 & G_\mathcal{C} = 1 - \frac{1}{\mathcal{F}_2 V} \frac{\mathcal{F}_4}{8 \mathcal{F}_2} + \frac{1}{(\mathcal{F}_2 V)^2} \bigg(-\frac{\mathcal{F}_6}{48 \mathcal{F}_2} + \frac{35 \mathcal{F}_4^2}{384 \mathcal{F}_2^2}\bigg) \\
 & \hspace{1.4cm} + \frac{1}{(\mathcal{F}_2 V)^3} \bigg(-\frac{\mathcal{F}_8}{384 \mathcal{F}_2} + \frac{7 \mathcal{F}_4 \mathcal{F}_6}{128 \mathcal{F}_2^2} - \frac{385 \mathcal{F}_4^3}{3072 \mathcal{F}_2^3} + \bigg(\frac{\mathcal{F}_6}{16 \mathcal{F}_2} - \frac{\mathcal{F}_4^2}{3 \mathcal{F}_2^2}\bigg) Q^2\bigg) \\
 & \hspace{1.4cm} + \mathcal{O}\bigg(\frac{1}{(\mathcal{E}_2 V)^4} \ , \ \frac{1}{(\mathcal{E}_2 V)^4} Q^2\bigg) = \\
 & \hspace{0.7cm} = 1 - \frac{1}{\mathcal{E}_2 V} \frac{\mathcal{E}_4 (1 + x_4 / \mathcal{E}_4 V)}{8 \mathcal{E}_2 (1 + x_2 / \mathcal{E}_2 V)^2} \\
 & \hspace{1.4cm} + \frac{1}{(\mathcal{E}_2 V)^2} \bigg(-\frac{\mathcal{E}_6 (1 + x_6 / \mathcal{E}_6 V)}{48 \mathcal{E}_2 (1 + x_2 / \mathcal{E}_2 V)^3} + \frac{35 \mathcal{E}_4^2 (1 + x_4 / \mathcal{E}_4 V)^2}{384 \mathcal{E}_2^2 (1 + x_2 / \mathcal{E}_2 V)^4}\bigg) \\
 & \hspace{1.4cm} + \frac{1}{(\mathcal{E}_2 V)^3} \bigg(-\frac{\mathcal{E}_8 (1 + x_8 / \mathcal{E}_8 V)}{384 \mathcal{E}_2 (1 + x_2 / \mathcal{E}_2 V)^4} + \frac{7 \mathcal{E}_4 (1 + x_4 / \mathcal{E}_4 V) \mathcal{E}_6 (1 + x_6 / \mathcal{E}_6 V)}{128 \mathcal{E}_2^2 (1 + x_2 / \mathcal{E}_2 V)^5} \\
 & \hspace{2.1cm} - \frac{385 \mathcal{E}_4^3 (1 + x_4 / \mathcal{E}_4 V)^3}{3072 \mathcal{E}_2^3 (1 + x_2 / \mathcal{E}_2 V)^6} + \bigg(\frac{\mathcal{E}_6 (1 + x_6 / \mathcal{E}_6 V)}{16 \mathcal{E}_2 (1 + x_2 / \mathcal{E}_2 V)^4} - \frac{\mathcal{E}_4^2 (1 + x_4 / \mathcal{E}_4 V)^2}{3 \mathcal{E}_2^2 (1 + x_2 / \mathcal{E}_2 V)^5}\bigg) Q^2\bigg) \\
 & \hspace{1.4cm} + \mathcal{O}\bigg(\frac{1}{(\mathcal{E}_2 V)^4} \ , \ \frac{1}{(\mathcal{E}_2 V)^4} Q^2\bigg) .
\end{aligned}
\end{equation}

After inserting $Z_{Q,V}$ (eq.\ (\ref{EQN010})), it is straightforward to obtain the final result for two-point correlation functions at fixed topology,
\begin{equation}
\label{EQN673} \begin{aligned}
 & C_{Q,V}(t) = \frac{\alpha(0)}{\sqrt{1 + x_2 / \mathcal{E}_2 V}} \\
 & \hspace{1.4cm} \exp\bigg(-M_H(0) t - \frac{1}{\mathcal{E}_2 V} \bigg(\frac{1}{1 + x_2 / \mathcal{E}_2 V} - 1\bigg) \frac{1}{2} Q^2 \\
 & \hspace{2.1cm} - \frac{1}{(\mathcal{E}_2 V)^3} \frac{\mathcal{E}_4 }{24 \mathcal{E}_2} \bigg(\frac{1 + x_4 / \mathcal{E}_4 V}{(1 + x_2 / \mathcal{E}_2 V)^4} -1\bigg) Q^4\bigg) \\
 & \hspace{1.4cm} \bigg(1 - \frac{1}{(\mathcal{E}_2 V)^2} \frac{\mathcal{E}_4}{2 \mathcal{E}_2} Q^2\bigg)^{+1/2} \bigg(1 - \frac{1}{(\mathcal{E}_2 V)^2} \frac{\mathcal{E}_4 (1 + x_4 / \mathcal{E}_4 V)}{2 \mathcal{E}_2 (1 + x_2 / \mathcal{E}_2 V)^3} Q^2\bigg)^{-1/2} \frac{G_\mathcal{C}}{G} \\
 & \hspace{0.7cm} + \mathcal{O}\bigg(\frac{1}{(\mathcal{E}_2 V)^4} Q^4\bigg) ,
\end{aligned}
\end{equation}
where $G$ and $G_\mathcal{C}$ are given in (\ref{EQN704}) and (\ref{EQN387}) (note that after inserting $G$ and $G_\mathcal{C}$ in (\ref{EQN673}) the error is $\mathcal{O}( 1 / (\mathcal{E}_2 V)^4 \ , \ Q^2 / (\mathcal{E}_2 V)^4 \ , \ Q^4 / (\mathcal{E}_2 V)^4)$). For some applications it might be helpful to have an expression for two-point correlation functions at fixed topology, which is of the form
\begin{equation}
\label{EQN516} C_{Q,V}(t) = \textrm{const} \times \exp\Big(-M_H(0) t + \textrm{fixed topology corrections as a power series in }1/\mathcal{E}_2 V\Big) ,
\end{equation}
i.e.\ where fixed topology effects only appear in the exponent and are sorted according to powers of $1/\mathcal{E}_2 V$. Such an expression can be obtained in a straightforward way from (\ref{EQN673}),
\begin{equation}
\label{EQN674} \begin{aligned}
 & C_{Q,V}(t) = \alpha(0) \exp\bigg(-M_H(0) t - \frac{1}{\mathcal{E}_2 V} \frac{x_2}{2} - \frac{1}{(\mathcal{E}_2 V)^2} \bigg(\frac{x_4 - 2 (\mathcal{E}_4/\mathcal{E}_2) x_2 - 2 x_2^2}{8} - \frac{x_2}{2} Q^2\bigg) \\
 & \hspace{1.4cm} - \frac{1}{(\mathcal{E}_2 V)^3} \\
 & \hspace{2.1cm} \bigg(\frac{16 (\mathcal{E}_4/\mathcal{E}_2)^2 x_2 + x_6 - 3 (\mathcal{E}_6/\mathcal{E}_2) x_2 - 8 (\mathcal{E}_4/\mathcal{E}_2) x_4 - 12 x_2 x_4 + 18 (\mathcal{E}_4/\mathcal{E}_2) x_2^2 + 8 x_2^3}{48} \\
 & \hspace{2.8cm} - \frac{x_4 - 3 (\mathcal{E}_4/\mathcal{E}_2) x_2 - 2 x_2^2}{4} Q^2\bigg)\bigg) \\
 & \hspace{0.7cm} + \mathcal{O}\bigg(\frac{1}{(\mathcal{E}_2 V)^4} \ , \ \frac{1}{(\mathcal{E}_2 V)^4} Q^2 \ , \ \frac{1}{(\mathcal{E}_2 V)^4} Q^4\bigg) .
\end{aligned}
\end{equation}
Note that the order of the error is the same for both (\ref{EQN673}) and (\ref{EQN674}).


\subsection{Comparison with \cite{Brower:2003yx}}

One can see that equations (3.8) and (3.9) derived in \cite{Brower:2003yx},
\begin{equation}
\label{EQN708} \Big\langle O(t_1) O(t_2) \Big\rangle \sim A_Q e^{-M_Q (t_1 - t_2)}
\end{equation}
and
\begin{equation}
\label{EQN709} M_Q = M(0) + \frac{1}{2} M''(0) \frac{1}{\beta V \chi_t} \bigg(1 - \frac{Q^2}{\beta V \chi_t}\bigg) + \ldots
\end{equation}
are contained in our result (\ref{EQN673}) and (\ref{EQN674}), respectively, after changing notation according to $\langle O(t_1) O(t_2) \rangle \rightarrow C_{Q,V}(t)$, $A_Q \rightarrow \alpha(0)$, $M(0) \rightarrow M_H(0)$ and $t_1 - t_2 \rightarrow t$.


\subsection{\label{SEC692}Parity mixing}

Parity $P$ is not a symmetry at $\theta \neq 0$. Therefore, states at $\theta \neq 0$ cannot be classified according to parity and it is not possible to construct two-point correlation functions $\mathcal{C}_{\theta,V}(t)$, where only $P = -$ or $P = +$ states contribute. Similarly, $C_{Q,V}(t)$ contains contributions both of states with $P = -$ or $P = +$, since it is obtained by Fourier transforming $\mathcal{C}_{\theta,V}(t)$ (cf.\ (\ref{EQN415})). Consequently, one has to determine the masses of $P = -$ and $P = +$ parity partners from the same two-point correlation functions\footnote{Note the similarity to twisted mass lattice QCD, where parity is also not an exact symmetry, and where $P = -$ and $P = +$ states are usually extracted from the same correlation matrix (cf.\ e.g.\ \cite{Jansen:2008si,Michael:2010aa,Baron:2010th,Wagner:2011fs,Kalinowski:2012re,Kalinowski:2013wsa,Wagner:2013laa}).}. While usually there are little problems for the lighter state (in the case of mesons typically the $P = -$ ground state), its parity partner (the $P = +$ ground state) has to be treated as an excitation. To precisely determine the mass of an excited state, a single correlator is in most cases not sufficient. For example to extract a first excitation it is common to study at least a $2 \times 2$ correlation matrix formed by two hadron creation operators, which generate significant overlap to both the ground state and the first excitation.

We discuss the determination of $P = -$ and $P = +$ parity partners from fixed topology computations in a simple setup: a $2 \times 2$ correlation matrix
\begin{equation}
C_{Q,V}^{j k}(t) \equiv \frac{1}{Z_{Q,V}} \int DA \, D\psi \, D\bar{\psi} \, \delta_{Q,Q[A]} O_j^\dagger(t) O_k(0) e^{-S_E[A,\bar{\psi},\psi]}
\end{equation}
with hadron creation operators $O_-$ and $O_+$ generating at unfixed topology and small $\theta$ mainly $P = -$ and $P = +$, respectively. An example for such operators is
\begin{equation}
\label{EQN508} O_- \equiv \frac{1}{\sqrt{V_s}} \int d^3r \, \bar{c}(\mathbf{r}) \gamma_5 u(\mathbf{r}) \quad , \quad O_+ \equiv \frac{1}{\sqrt{V_s}} \int d^3r \, \bar{c}(\mathbf{r}) u(\mathbf{r})
\end{equation}
corresponding to the $D$ mesons and its parity partner $D_0^\ast$. Without loss of generality we assume that the ground state (at $\theta = 0$) has $P = -$, denoted by $H_-$, and the first excitation has $P = +$, denoted by $H_+$.

In the following we derive expressions for the four elements of the correlation matrix $C_{Q,V}^{j k}(t)$, $j,k \in \{ - , + \}$. We proceed similar as in section~\ref{SEC346}. This time, however, we consider the two lowest states $H_-$ and $H_+$ (not only a single state),
\begin{equation}
\label{EQN755} \mathcal{C}_{\theta,V}^{j k}(t) \mathcal{Z}_{\theta,V} = \Big(\alpha_-^{j k}(\theta,V_s) e^{-M_{H_-}(\theta) t} + \alpha_+^{j k}(\theta,V_s) e^{-M_{H_+}(\theta) t}\Big) e^{-E_0(\theta,V_s) T}
\end{equation}
(which is the generalization of (\ref{EQN856}) with exponentially suppressed corrections from higher excitations neglected), where
\begin{equation}
\label{EQN461} \alpha_n^{j k}(\theta) \equiv A_n^{j,\dagger}(\theta) A_n^k(\theta) \quad , \quad A_n^j(\theta) \equiv \langle H_n ; \theta | O_j | 0 ; \theta \rangle .
\end{equation}

The overlaps of the trial states $O_j | 0 ; \theta \rangle$ and the lowest states $| H_n \rangle$, $A_n^j(\theta)$ and $\alpha_n^{j k}(\theta)$, have to be treated in a more general way, since the leading order of their $\theta$ expansion can be proportional to a constant, to $\theta$ or to $\theta^2$ depending on the indices $j$, $k$ and $n$. Since at $\theta = 0$ parity is a symmetry, $A_-^+(\theta = 0) = A_+^-(\theta = 0) = 0$. Consequently,
\begin{itemize}
\item $A_-^+(\theta) = \mathcal{O}(\theta)$, $A_+^-(\theta) = \mathcal{O}(\theta)$,
\end{itemize}
while
\begin{itemize}
\item $A_-^-(\theta) = \mathcal{O}(1)$, $A_+^+(\theta) = \mathcal{O}(1)$.
\end{itemize}
From the definition of $\alpha_n^{j k}(\theta)$ (eq.\ (\ref{EQN461})) one can conclude
\begin{itemize}
\item $\alpha_-^{- -}(\theta) = \mathcal{O}(1)$, $\alpha_+^{+ +}(\theta) = \mathcal{O}(1)$, 

\item $\alpha_\pm^{- +}(\theta) = \mathcal{O}(\theta)$, $\alpha_\pm^{+ -}(\theta) = \mathcal{O}(\theta)$, 

\item $\alpha_-^{+ +}(\theta) = \mathcal{O}(\theta^2)$, $\alpha_+^{- -}(\theta) = \mathcal{O}(\theta^2)$.
\end{itemize}
Using $P O_\pm P = \pm O_\pm$ and $P | n ; +\theta , V_s \rangle = \eta_n(\theta , V_s) | n ; -\theta , V_s \rangle$, where $\eta_n(\theta , V_s)$ is a non-unique phase, one can show
\begin{itemize}
\item $\alpha_n^{+ +}(+\theta) = +\alpha_n^{+ +}(-\theta)$, $\alpha_n^{- -}(+\theta) = +\alpha_n^{- -}(-\theta)$ (i.e.\ only even powers of $\theta$ in the corresponding expansions),

\item $\alpha_n^{+ -}(+\theta) = -\alpha_n^{+ -}(-\theta)$, $\alpha_n^{- +}(+\theta) = -\alpha_n^{- +}(-\theta)$ (i.e.\ only odd powers of $\theta$ in the corresponding expansions).
\end{itemize}

Technically it is straightforward to consider not only the ground state $H_-$, but also a first excitation $H_+$: the contributions of the two states are just summed in (\ref{EQN755}), i.e.\ one can independently determine their Fourier transform and, hence, their contribution to the correlation matrix at fixed topology, $C_{Q,V}^{j k}(t)$. Additional calculations have to be done, however, for off-diagonal elements, where $\alpha_n^{\pm \mp}(+\theta) = -\alpha_n^{\pm \mp}(-\theta)$, and for contributions to diagonal matrix elements, where $\alpha_n^{\pm \pm}(\theta) = \mathcal{O}(\theta^2)$ (cf.\ the following two subsections). Contributions to diagonal matrix elements, where $\alpha_n^{\pm \pm}(\theta) = \mathcal{O}(1)$, have already been determined (cf.\ section~\ref{SEC346}).


\subsubsection{Calculation for $\alpha(+\theta) = -\alpha(-\theta)$, where $\alpha(\theta) \in \{ \alpha_-^{- +}(\theta), \alpha_+^{- +}(\theta), \alpha_-^{+ -}(\theta), \alpha_+^{+ -}(\theta) \}$}

We proceed as in section~\ref{SEC346}. $\alpha(\theta)$ can be written as a power series around $\theta = 0$,
\begin{equation}
\alpha(\theta) = \sum_{k=0}^\infty \frac{\alpha^{(2k+1)}(0) \theta^{2k+1}}{(2k+1)!} = \alpha^{(1)}(0) \theta \exp\bigg(\underbrace{\ln\bigg(\sum_{k=0}^\infty \frac{\alpha^{(2k+1)}(0) \theta^{2k}}{(2k+1)! \alpha^{(1)}(0)}\bigg)}_{\equiv -\beta(\theta) = -\sum_{k=1}^\infty \frac{\beta^{(2k)}(0) \theta^{2k}}{(2k)!}}\bigg) .
\end{equation}
The corresponding contribution to $\mathcal{C}_{\theta,V}^{j k}(t) \mathcal{Z}_{\theta,V}$ (cf.\ (\ref{EQN755})) is
\begin{equation}
\alpha(\theta) e^{-M_H(\theta) t} e^{-E_0(\theta,V_s) T} = \alpha^{(1)}(0) \theta e^{-(e_0(\theta) V + M_H(\theta) t + \beta(\theta))} .
\end{equation}
As before we define
\begin{equation}
f_{\mathcal{C}}(\theta) \equiv f_{\mathcal{C}}(\theta,Q,V) \equiv e_0(\theta) + \frac{M_H(\theta) t + \beta(\theta) - i Q \theta}{V} .
\end{equation}
For the contribution to the correlation matrix at fixed topology $C_{Q,V}^{j k}(t) Z_{Q,V}$ we then obtain
\begin{equation}
\label{EQN239} \frac{\alpha^{(1)}(0)}{2 \pi} \int_{-\pi}^{+\pi} d\theta \, \theta e^{-f_{\mathcal{C}}(\theta) V} ,
\end{equation}
where $f_{\mathcal{C}}(\theta) V$ is defined by (\ref{EQN995}) and (\ref{EQN079}). Consequently, its minimum $\theta_{s,\mathcal{C}}$ is given by (\ref{EQN357}). (\ref{EQN239}) can be written as
\begin{equation}
\label{EQN764} \begin{aligned}
 & \frac{\alpha^{(1)}(0) e^{-f_{\mathcal{C}}(\theta_{s,\mathcal{C}}) V}}{(2 \pi f_{\mathcal{C}}^{(2)}(\theta_{s,\mathcal{C}}) V)^{1/2}} \bigg\Vert \bigg(\theta_{s,\mathcal{C}} + \frac{s}{(f_{\mathcal{C}}^{(2)}(\theta_{s,\mathcal{C}}) V)^{1/2}}\bigg) \exp\bigg(-\sum_{n=3}^\infty \frac{f_{\mathcal{C}}^{(n)}(\theta_{s,\mathcal{C}}) V}{n! (f_{\mathcal{C}}^{(2)}(\theta_{s,\mathcal{C}}) V)^{n/2}} s^n\bigg)\bigg\Vert = \\
 & \hspace{0.7cm} = \frac{\alpha^{(1)}(0) e^{-f_{\mathcal{C}}(\theta_{s,\mathcal{C}}) V}}{(2 \pi f_{\mathcal{C}}^{(2)}(\theta_{s,\mathcal{C}}) V)^{1/2}} \Big(\theta_{s,\mathcal{C}} G_\mathcal{C} + H_\mathcal{C}\Big) ,
\end{aligned}
\end{equation}
where $G_\mathcal{C}$ is defined in (\ref{eq:saddle point C_Q}) and
\begin{equation}
\label{EQN883} \begin{aligned}
 & H_\mathcal{C} \equiv \bigg\Vert \frac{s}{(f_{\mathcal{C}}^{(2)}(\theta_{s,\mathcal{C}}) V)^{1/2}} \exp\bigg(-\sum_{n=3}^\infty \frac{f_{\mathcal{C}}^{(n)}(\theta_{s,\mathcal{C}}) V}{n! (f_{\mathcal{C}}^{(2)}(\theta_{s,\mathcal{C}}) V)^{n/2}} s^n\bigg)\bigg\Vert = \\
 & \hspace{0.7cm} = \sum_{k=1}^\infty \frac{(-1)^k}{k!} \bigg\Vert \frac{s}{(f_{\mathcal{C}}^{(2)}(\theta_{s,\mathcal{C}}) V)^{1/2}} \bigg(\sum_{n=3}^\infty \frac{f_{\mathcal{C}}^{(n)}(\theta_{s,\mathcal{C}}) V}{n! (f_{\mathcal{C}}^{(2)}(\theta_{s,\mathcal{C}}) V)^{n/2}} s^n\bigg)^k\bigg\Vert .
\end{aligned}
\end{equation}
As in section~\ref{SEC345} it is easy to identify and calculate all terms of $H_\mathcal{C}$ up to $\mathcal{O}(1/(\mathcal{E}_2 V)^3)$:
\begin{itemize}
\item $k=1$, $n=3$ ($\propto 1/V^2$):
\begin{equation}
\bigg\Vert \frac{f_{\mathcal{C}}^{(3)}(\theta_{s,\mathcal{C}}) V}{3! (f_{\mathcal{C}}^{(2)}(\theta_{s,\mathcal{C}}) V)^2} s^4\bigg\Vert = 
\frac{1}{\mathcal{F}_2 V} \frac{\mathcal{F}_4}{2 \mathcal{F}_2} \theta_{s,\mathcal{C}}
+ \mathcal{O}\bigg(\frac{1}{(\mathcal{E}_2 V)^4}\bigg) .
\end{equation}

\item $k=1$, $n=5$ ($\propto 1/V^3$):
\begin{equation}
\bigg\Vert \frac{f_{\mathcal{C}}^{(5)}(\theta_{s,\mathcal{C}}) V}{5! (f_{\mathcal{C}}^{(2)}(\theta_{s,\mathcal{C}}) V)^3} s^6\bigg\Vert =
\frac{1}{(\mathcal{F}_2 V)^2} \frac{\mathcal{F}_6}{8 \mathcal{F}_2} \theta_{s,\mathcal{C}}
+ \mathcal{O}\bigg(\frac{1}{(\mathcal{E}_2 V)^4}\bigg) .
\end{equation}

\item $k=2$, $n_1=3$, $n_2=4$ ($\propto 1/V^3$):
\begin{equation}
2 \times \bigg\Vert \frac{(f_{\mathcal{C}}^{(3)}(\theta_{s,\mathcal{C}}) V) (f_{\mathcal{C}}^{(4)}(\theta_{s,\mathcal{C}}) V)}{3! 4! (f_{\mathcal{C}}^{(2)}(\theta_{s,\mathcal{C}}) V)^4} s^8\bigg\Vert =
\frac{1}{(\mathcal{F}_2 V)^2} \frac{35 \mathcal{F}_4^2}{24 \mathcal{F}_2^2} \theta_{s,\mathcal{C}}
+ \mathcal{O}\bigg(\frac{1}{(\mathcal{E}_2 V)^4}\bigg) .
\end{equation}
\end{itemize}
Inserting these expressions into (\ref{EQN883}) leads to
\begin{equation}
H_\mathcal{C} =
-\frac{1}{\mathcal{F}_2 V} \frac{\mathcal{F}_4}{2 \mathcal{F}_2} \theta_{s,\mathcal{C}}
+ \frac{1}{(\mathcal{F}_2 V)^2} \bigg(-\frac{\mathcal{F}_6}{8 \mathcal{F}_2} + \frac{35 \mathcal{F}_4^2}{48 \mathcal{F}_2^2}\bigg) \theta_{s,\mathcal{C}}
+ \mathcal{O}\bigg(\frac{1}{(\mathcal{E}_2 V)^4}\bigg) .
\end{equation}
The final explicit expression up to $\mathcal{O}(1 / (\mathcal{E}_2 V)^3)$ for the contribution to $C_{Q,V}^{j k}(t) Z_{Q,V}$ (eq.\ (\ref{EQN764})) is
\begin{equation}
\begin{aligned}
 & \frac{\alpha^{(1)}(0)}{\sqrt{2 \pi \mathcal{F}_2 V}} \bigg(\exp\bigg(-\mathcal{F}_0 V - \frac{1}{\mathcal{F}_2 V} \frac{1}{2} Q^2 - \frac{1}{(\mathcal{F}_2 V)^3} \frac{\mathcal{F}_{4}}{24 \mathcal{F}_2} Q^4\bigg) \\
 & \hspace{2.1cm} \bigg(1 - \frac{1}{(\mathcal{F}_2 V)^2} \frac{\mathcal{F}_4}{2 \mathcal{F}_2} Q^2\bigg)^{-1/2} \Big(\theta_{s,\mathcal{C}} G_\mathcal{C} + H_\mathcal{C}\Big) + \mathcal{O}\bigg(\frac{1}{(\mathcal{E}_2 V)^4} Q^4\bigg)\bigg) .
\end{aligned}
\end{equation}
After dividing by $Z_{Q,V}$ (eq.\ (\ref{EQN010})), it is straightforward to obtain the final result. In exponential form (\ref{EQN516}) it is
\begin{equation}
\label{EQN997} \begin{aligned}
 & C_{Q,V}^{j k}(t) \leftarrow \frac{i \alpha^{(1)}(0) Q}{\mathcal{E}_2 V} \exp\bigg(-M_H(0) t 
- \frac{1}{\mathcal{E}_2 V} \bigg(\frac{(\mathcal{E}_4/\mathcal{E}_2) + 3 x_2}{2}\bigg) \\
 & \hspace{1.4cm} - \frac{1}{(\mathcal{E}_2 V)^2} \bigg(
\frac{
3 (\mathcal{E}_6/\mathcal{E}_2)
- 13 (\mathcal{E}_4/\mathcal{E}_2)^2
- 30 (\mathcal{E}_4/\mathcal{E}_2) x_2
+ 15 x_4
- 18 x_2^2
}{24} \\
 & \hspace{2.8cm} - \frac{
(\mathcal{E}_4/\mathcal{E}_2)
+ 3 x_2
}{6} Q^2
\bigg) \\
 & \hspace{0.7cm} + \mathcal{O}\bigg(\frac{1}{(\mathcal{E}_2 V)^4}\bigg) .
\end{aligned}
\end{equation}


\subsubsection{Calculation for $\alpha(+\theta) = +\alpha(-\theta)$, where $\alpha(\theta) \in \{ \alpha_-^{+ +}(\theta), \alpha_+^{- -}(\theta) \} = \mathcal{O}(\theta^2)$}

We proceed as in section~\ref{SEC346}. $\alpha(\theta)$ can be written as a power series around $\theta = 0$,
\begin{equation}
\alpha(\theta) = \sum_{k=1}^\infty \frac{\alpha^{(2k)}(0) \theta^{2k}}{(2k)!} = \frac{\alpha^{(2)}(0)}{2} \theta^2 \exp\bigg(\underbrace{\ln\bigg(\sum_{k=0}^\infty \frac{2 \alpha^{(2k+2)}(0) \theta^{2k}}{(2k+2)! \alpha^{(2)}(0)}\bigg)}_{\equiv -\beta(\theta) = -\sum_{k=1}^\infty \frac{\beta^{(2k)}(0) \theta^{2k}}{(2k)!}}\bigg) .
\end{equation}
The corresponding contribution to $\mathcal{C}_{\theta,V}^{j k}(t) \mathcal{Z}_{\theta,V}$ (cf.\ (\ref{EQN755})) is
\begin{equation}
\alpha(\theta) e^{-M_H(\theta) t} e^{-E_0(\theta,V_s) T} = \frac{\alpha^{(2)}(0)}{2} \theta^2 e^{-(e_0(\theta) V + M_H(\theta) t + \beta(\theta))} .
\end{equation}
As before we define
\begin{equation}
f_{\mathcal{C}}(\theta) \equiv f_{\mathcal{C}}(\theta,Q,V) \equiv e_0(\theta) + \frac{M_H(\theta) t + \beta(\theta) - i Q \theta}{V} .
\end{equation}
For the contribution to the correlation matrix at fixed topology $C_{Q,V}^{j k}(t) Z_{Q,V}$ we then obtain
\begin{equation}
\label{EQN239_} \frac{\alpha^{(2)}(0)}{4 \pi} \int_{-\pi}^{+\pi} d\theta \, \theta^2 e^{-f_{\mathcal{C}}(\theta) V} ,
\end{equation}
where $f_{\mathcal{C}}(\theta) V$ is defined by (\ref{EQN995}) and (\ref{EQN079}). Consequently, its minimum $\theta_{s,\mathcal{C}}$ is given by (\ref{EQN357}). (\ref{EQN239_}) can be written as
\begin{equation}
\label{EQN764_} \begin{aligned}
 & \frac{\alpha^{(2)}(0) e^{-f_{\mathcal{C}}(\theta_{s,\mathcal{C}}) V}}{2 (2 \pi f_{\mathcal{C}}^{(2)}(\theta_{s,\mathcal{C}}) V)^{1/2}} \bigg\Vert \bigg(\theta_{s,\mathcal{C}} + \frac{s}{(f_{\mathcal{C}}^{(2)}(\theta_{s,\mathcal{C}}) V)^{1/2}}\bigg)^2 \exp\bigg(-\sum_{n=3}^\infty \frac{f_{\mathcal{C}}^{(n)}(\theta_{s,\mathcal{C}}) V}{n! (f_{\mathcal{C}}^{(2)}(\theta_{s,\mathcal{C}}) V)^{n/2}} s^n\bigg)\bigg\Vert = \\
 & \hspace{0.7cm} = \frac{\alpha^{(2)}(0) e^{-f_{\mathcal{C}}(\theta_{s,\mathcal{C}}) V}}{2 (2 \pi f_{\mathcal{C}}^{(2)}(\theta_{s,\mathcal{C}}) V)^{1/2}} \Big(\theta_{s,\mathcal{C}}^2 G_\mathcal{C} + 2 \theta_{s,\mathcal{C}} H_\mathcal{C} + I_\mathcal{C}\Big) ,
\end{aligned}
\end{equation}
where $G_\mathcal{C}$ is defined in (\ref{eq:saddle point C_Q}), $H_\mathcal{C}$ is defined in (\ref{EQN883}) and
\begin{equation}
\label{EQN883_} \begin{aligned}
 & I_\mathcal{C} \equiv \bigg\Vert \frac{s^2}{f_{\mathcal{C}}^{(2)}(\theta_{s,\mathcal{C}}) V} \exp\bigg(-\sum_{n=3}^\infty \frac{f_{\mathcal{C}}^{(n)}(\theta_{s,\mathcal{C}}) V}{n! (f_{\mathcal{C}}^{(2)}(\theta_{s,\mathcal{C}}) V)^{n/2}} s^n\bigg)\bigg\Vert = \\
 & \hspace{0.7cm} = \sum_{k=0}^\infty \frac{(-1)^k}{k!} \bigg\Vert \frac{s^2}{f_{\mathcal{C}}^{(2)}(\theta_{s,\mathcal{C}}) V} \bigg(\sum_{n=3}^\infty \frac{f_{\mathcal{C}}^{(n)}(\theta_{s,\mathcal{C}}) V}{n! (f_{\mathcal{C}}^{(2)}(\theta_{s,\mathcal{C}}) V)^{n/2}} s^n\bigg)^k\bigg\Vert .
\end{aligned}
\end{equation}
As in section~\ref{SEC345} it is easy to identify and calculate all terms of $I_\mathcal{C}$ up to $\mathcal{O}(1/(\mathcal{E}_2 V)^3)$:
\begin{itemize}
\item $k=0$ ($\propto 1/V$):
\begin{equation}
\bigg\Vert \frac{s^2}{f_{\mathcal{C}}^{(2)}(\theta_{s,\mathcal{C}}) V}\bigg\Vert =
\frac{1}{\mathcal{F}_2 V}
- \frac{1}{\mathcal{F}_2 V} \frac{\mathcal{F}_4}{2 \mathcal{F}_2} \theta_{s,\mathcal{C}}^2
+ \mathcal{O}\bigg(\frac{1}{(\mathcal{E}_2 V)^4}\bigg) .
\end{equation}

\item $k=1$, $n=4$ ($\propto 1/V^2$):
\begin{equation}
\bigg\Vert \frac{f_{\mathcal{C}}^{(4)}(\theta_{s,\mathcal{C}}) V}{4! (f_{\mathcal{C}}^{(2)}(\theta_{s,\mathcal{C}}) V)^3} s^6\bigg\Vert =
\frac{1}{(\mathcal{F}_2 V)^2} \frac{5 \mathcal{F}_4}{8 \mathcal{F}_2}
+ \mathcal{O}\bigg(\frac{1}{(\mathcal{E}_2 V)^4}\bigg) .
\end{equation}

\item $k=1$, $n=6$ ($\propto 1/V^3$):
\begin{equation}
\bigg\Vert \frac{f_{\mathcal{C}}^{(6)}(\theta_{s,\mathcal{C}}) V}{6! (f_{\mathcal{C}}^{(2)}(\theta_{s,\mathcal{C}}) V)^4} s^8\bigg\Vert =
\frac{1}{(\mathcal{F}_2 V)^3} \frac{7 \mathcal{F}_6}{48 \mathcal{F}_2}
+ \mathcal{O}\bigg(\frac{1}{(\mathcal{E}_2 V)^4}\bigg) .
\end{equation}

\item $k=2$, $n_1=4$, $n_2=4$ ($\propto 1/V^3$):
\begin{equation}
\bigg\Vert \frac{(f_{\mathcal{C}}^{(4)}(\theta_{s,\mathcal{C}}) V)^2}{(4!)^2 (f_{\mathcal{C}}^{(2)}(\theta_{s,\mathcal{C}}) V)^5} s^{10}\bigg\Vert =
\frac{1}{(\mathcal{F}_2 V)^3} \frac{105 \mathcal{F}_4^2}{64 \mathcal{F}_2^2}
+ \mathcal{O}\bigg(\frac{1}{(\mathcal{E}_2 V)^4}\bigg) .
\end{equation}
\end{itemize}
Inserting these expressions into (\ref{EQN883_}) leads to
\begin{equation}
I_\mathcal{C} =
\frac{1}{\mathcal{F}_2 V} \bigg(1 - \frac{\mathcal{F}_4}{2 \mathcal{F}_2} \theta_{s,\mathcal{C}}^2\bigg)
+ \frac{1}{(\mathcal{F}_2 V)^2} \bigg(-\frac{5 \mathcal{F}_4}{8 \mathcal{F}_2}\bigg)
+ \frac{1}{(\mathcal{F}_2 V)^3} \bigg(-\frac{7 \mathcal{F}_6}{48 \mathcal{F}_2} + \frac{105 \mathcal{F}_4^2}{128 \mathcal{F}_2^2}\bigg)
+ \mathcal{O}\bigg(\frac{1}{(\mathcal{E}_2 V)^4}\bigg) .
\end{equation}

The final explicit expression up to $\mathcal{O}(1 / (\mathcal{E}_2 V)^3)$ for the contribution to $C_{Q,V}^{j k}(t) Z_{Q,V}$ (eq.\ (\ref{EQN764_})) is
\begin{equation}
\begin{aligned}
 & \frac{\alpha^{(2)}(0)}{2 \sqrt{2 \pi \mathcal{F}_2 V}} \bigg(\exp\bigg(-\mathcal{F}_0 V - \frac{1}{\mathcal{F}_2 V} \frac{1}{2} Q^2 - \frac{1}{(\mathcal{F}_2 V)^3} \frac{\mathcal{F}_{4}}{24 \mathcal{F}_2} Q^4\bigg) \\
 & \hspace{2.1cm} \bigg(1 - \frac{1}{(\mathcal{F}_2 V)^2} \frac{\mathcal{F}_4}{2 \mathcal{F}_2} Q^2\bigg)^{-1/2} \Big(\theta_{s,\mathcal{C}}^2 G_\mathcal{C} + 2 \theta_{s,\mathcal{C}} H_\mathcal{C} + I_\mathcal{C}\Big) + \mathcal{O}\bigg(\frac{1}{(\mathcal{E}_2 V)^4} Q^4\bigg)\bigg) .
\end{aligned}
\end{equation}
After dividing by $Z_{Q,V}$ (eq.\ (\ref{EQN010})), it is straightforward to obtain the final result. In exponential form (\ref{EQN516}) it is
\begin{equation}
\label{EQN996} \begin{aligned}
 & C_{Q,V}^{j k}(t) \leftarrow \frac{\alpha^{(2)}(0)}{2 \mathcal{E}_2 V} \exp\bigg(-M_H(0) t 
- \frac{1}{\mathcal{E}_2 V} \bigg(\frac{(\mathcal{E}_4/\mathcal{E}_2) + 3 x_2}{2} + Q^2\bigg) \\
 & \hspace{1.4cm} - \frac{1}{(\mathcal{E}_2 V)^2} \bigg(
\frac{
3 (\mathcal{E}_6/\mathcal{E}_2)
- 13 (\mathcal{E}_4/\mathcal{E}_2)^2
- 30 (\mathcal{E}_4/\mathcal{E}_2) x_2
+ 15 x_4
- 18 x_2^2
}{24} \\
 & \hspace{2.8cm} - \frac{
2 (\mathcal{E}_4/\mathcal{E}_2)
+ 3 x_2
}{2} Q^2 + \frac{1}{2} Q^4
\bigg) \\
 & \hspace{0.7cm} + \mathcal{O}\bigg(\frac{1}{(\mathcal{E}_2 V)^4}\bigg) .
\end{aligned}
\end{equation}


\subsubsection{The $2 \times 2$ correlation matrix at fixed topology at $\mathcal{O}(1 / \mathcal{E}_2 V)$}

The $2 \times 2$ correlation matrix $C_{Q,V}^{j k}(t)$, $j, k \in \{ - , + \}$ can be obtained by properly adding the results (\ref{EQN674}), (\ref{EQN997}) and (\ref{EQN996}). At first order in $1 / \mathcal{E}_2 V$ it is given by
\begin{eqnarray}
\nonumber & & \hspace{-0.7cm} C_{Q,V}^{--}(t) = \alpha_-^{--}(0) \exp\bigg(-M_{H_-}(0) t - \frac{1}{\mathcal{E}_2 V} \frac{x_{2,-}^{--}}{2}\bigg) + \frac{\alpha_+^{--,(2)}(0)}{2 \mathcal{E}_2 V} \exp\bigg(-M_{H_+}(0) t\bigg) \\
\label{EQN445} & & \hspace{0.675cm} + \mathcal{O}\bigg(\frac{1}{(\mathcal{E}_2 V)^2}\bigg) \\
\nonumber & & \hspace{-0.7cm} C_{Q,V}^{++}(t) = \alpha_+^{++}(0) \exp\bigg(-M_{H_+}(0) t - \frac{1}{\mathcal{E}_2 V} \frac{x_{2,+}^{++}}{2}\bigg) + \frac{\alpha_-^{++,(2)}(0)}{2 \mathcal{E}_2 V} \exp\bigg(-M_{H_-}(0) t\bigg) \\
\label{EQN447} & & \hspace{0.675cm} + \mathcal{O}\bigg(\frac{1}{(\mathcal{E}_2 V)^2}\bigg) \\
\nonumber & & \hspace{-0.7cm} C_{Q,V}^{\mp \pm}(t) = \frac{i \alpha_-^{\mp \pm,(1)}(0) Q}{\mathcal{E}_2 V} \exp\bigg(-M_{H_-}(0) t\bigg) + \frac{i \alpha_+^{\mp \pm,(1)}(0) Q}{\mathcal{E}_2 V} \exp\bigg(-M_{H_+}(0) t\bigg) \\
\label{EQN446} & & \hspace{0.675cm} + \mathcal{O}\bigg(\frac{1}{(\mathcal{E}_2 V)^2}\bigg) ,
\end{eqnarray}
where $x_{2,\pm}^{\pm \pm} = M_{H_\pm}^{(2)} t + \beta_\pm^{\pm \pm,(2)}$ and $\beta_\pm^{\pm \pm,(2)} = -\alpha_\pm^{\pm \pm,(2)}(0) / \alpha_\pm^{\pm \pm}(0)$ (cf.\ (\ref{EQN999})). The quantities $\alpha_n^{j k}$ are products of the more fundamental $A_n^j$ (cf.\ (\ref{EQN461})) and, therefore, are not independent and fulfill certain constraints. Since the diagonal elements of $\mathcal{C}_{\theta,V}^{j k}(t)$ are real and $\geq 0$,
\begin{itemize}
\item $\alpha_-^{--}(0) , \alpha_+^{++}(0) , \alpha_-^{++,(2)}(0) , \alpha_+^{--,(2)}(0) \geq 0$ and real (4 real parameters),

\item $\alpha_-^{--,(2)}(0) , \alpha_+^{++,(2)}(0)$ real (2 real parameters).
\end{itemize}
Moreover, from $(\mathcal{C}_{\theta,V}^{j k}(t))^\ast = \mathcal{C}_{\theta,V}^{k j}(t)$ follows
\begin{itemize}
\item $(\alpha_-^{-+,(1)}(0))^\ast = \alpha_-^{+-,(1)}(0)$ and $(\alpha_+^{-+,(1)}(0))^\ast = \alpha_+^{+-,(1)}(0)$ (4 real parameters).
\end{itemize}
Quite often one can define the hadron creation operators $O_-$ and $O_+$ in such a way that the off-diagonal elements of $\mathcal{C}_{\theta,V}^{j k}(t)$ are real (or purely imaginary), which reduces the number of real parameters contained in $\alpha_n^{j k}$ from 10 to 8. There are further parameters, $M_{H_-}(0)$, $M_{H_+}(0)$, $M_{H_-}^{(2)}(0)$, $M_{H_+}^{(2)}(0)$ and $\mathcal{E}_2$, i.e.\ in total 13 parameters.

(\ref{EQN445}) to (\ref{EQN446}) clearly show that parity mixing at fixed topology is already present at order $1 / \mathcal{E}_2 V$. In particular this will cause problems, when trying to extract a hadron, which has a lighter parity partner, from a single two-point correlation function: e.g.\ the first term in $C_{Q,V}^{++}(t)$ (eq.\ (\ref{EQN447})) is suited to determine a positive parity meson; however, there is a contamination by the corresponding lighter negative parity meson due to the second term, which is only suppressed proportional to $1 / \mathcal{E}_2 V$ with respect to the spacetime volume; since the first term is exponentially suppressed with respect to the temporal separation compared to the second term ($\propto e^{-(M_{H_+} - M_{H_-}) t}$), a precise determination of $M_{H_+}$ from the single correlator $C_{Q,V}^{++}(t)$ seems extremely difficult and would probably require extremely precise simulation results. Using the full $2 \times 2$ correlation matrix (\ref{EQN445}) to (\ref{EQN446}) should, however, stabilize a fit to extract $M_{H_+}$ and $M_{H_-}$ at the same time (this is discussed in detail in section~\ref{SEC439}), similar to what is usually done at ordinary unfixed topology computations, when determining excited states.

This parity mixing at fixed topology has already been observed and discussed in the context of the $\eta$ meson in \cite{Aoki:2007ka}. When considering the correlation function $C_{Q,V}^{--}(t)$ with a suitable $\eta$ meson creation operator, e.g.\
\begin{equation}
O_- \equiv \frac{1}{\sqrt{V_s}} \int d^3r \, \Big(\bar{u}(\mathbf{r}) \gamma_5 u(\mathbf{r}) + \bar{d}(\mathbf{r}) \gamma_5 d(\mathbf{r})\Big) ,
\end{equation}
one finds
\begin{equation}
C_{Q,V}^{--}(t) = \alpha_\eta^{--}(0) \exp\bigg(-M_\eta(0) t - \frac{1}{\mathcal{E}_2 V} \frac{x_{2,\eta}^{--}}{2}\bigg) + \frac{\alpha_0^{--,(2)}(0)}{2 \mathcal{E}_2 V} + \mathcal{O}\bigg(\frac{1}{(\mathcal{E}_2 V)^2}\bigg) ,
\end{equation}
where $M_{H_+} = 0$ has been used ($H_+$ is in this context the vacuum state). Using $\alpha_0^{--,(2)}(0) = -2 \mathcal{E}_2^2$ from \cite{Aoki:2007ka} shows that there is a time independent contribution $-\mathcal{E}_2 / V$ to the correlation function $C_{Q,V}^{--}(t)$ as in \cite{Aoki:2007ka}.

It is straightforward to extend (\ref{EQN445}) to (\ref{EQN446}) to larger correlation matrices formed by more than the two operators $O_-$ and $O_+$. Similarly, it is easy to include further states besides $H_-$ and $H_+$. In both cases one just has to properly add the expressions (\ref{EQN674}), (\ref{EQN997}) and (\ref{EQN996}) and assign suitable indices.


\section{\label{SEC867}Discussion of errors}

In this section we discuss, in which regime of parameters our $1/V$ expansions of two-point correlation functions at fixed topology (\ref{EQN673}) and (\ref{EQN674}) are accurate approximations.


\subsection{\label{EQN735}Errors proportional to $1 / \mathcal{E}_2 V$}

In section~\ref{SEC345} the spacetime dependence of two-point correlation functions $C_{Q,V}(t)$ has been derived up to $1/V^3$. More precisely, the error is
\begin{equation}
\mathcal{O}\bigg(\frac{1}{(\mathcal{E}_2 V)^4} \ , \ \frac{1}{(\mathcal{E}_2 V)^4} Q^2 \ , \ \frac{1}{(\mathcal{E}_2 V)^4} Q^4\bigg)
\end{equation}
(cf.\ (\ref{EQN673}), the text below (\ref{EQN673}) and (\ref{EQN674})). This error will be small, if
\begin{itemize}
\item[\textbf{(C1)}] $\phantom{xxx} 1 / \mathcal{E}_2 V \ll 1 \quad , \quad |Q| / \mathcal{E}_2 V \ll 1$.
\end{itemize}
In other words, computations at fixed topology require large spacetime volumes $V$ (in units of the topological susceptibility $\chi_t = \mathcal{E}_2$), while the topological charge $Q$ may not be too large. We have also used $\mathcal{F}_2 = \mathcal{E}_2 + \mathcal{O}(1 / \mathcal{E}_2 V)$, which requires 
\begin{itemize}
\item[\textbf{(C2)}] $\phantom{xxx} |x_2| = |M_H^{(2)}(0) t + \beta^{(2)}(0)| \ltapprox 1$.
\end{itemize}
The time dependence of this constraint excludes the use of large values of $t$.


\subsection{\label{SEC009}Exponentially suppressed errors}

In sections \ref{SEC345} and \ref{SEC346} several exponentially suppressed corrections have been neglected:
\begin{itemize}
\item[(a)] \textbf{Ordinary finite volume effects, i.e.\ finite volume effects not associated with fixed topology:} \\
Such finite volume effects also appear in QCD simulations, where topology is not fixed. These effects are expected to be proportional to $e^{-m_\pi(\theta) L}$, where $m_\pi(\theta)$ is the mass of the pion (the lightest hadron mass) and $L$ is the periodic spatial extension.

\item[(b)] \textbf{Contributions of excited states to the partition function and to two-point correlation functions:} \\
Excited states contribute to the partition function $\mathcal{Z}_{\theta,V}$ proportional to $e^{-\Delta E(\theta) T}$ (cf.\ (\ref{EQN003})), where $\Delta E(\theta) = E_1(\theta,V_s) - E_0(\theta,V_s)$ is the mass of the lightest hadron, i.e.\ $\Delta E(\theta) = m_\pi(\theta)$. \\
The corresponding dominating terms in a two-point correlation function $\mathcal{C}_{\theta,V}(t) \mathcal{Z}_{\theta,V}$ are proportional to $e^{-(M_H^\ast(\theta) - M_H(\theta)) t}$ and $e^{-M_H(\theta) (T-2 t)}$ (cf.\ (\ref{EQN856})), where $M_H(\theta)$ is the mass of the hadron of interest and $M_H^\ast(\theta) - M_H(\theta)$ the difference to its first excitation.

\item[(c)] \textbf{Changing the integration limits in (\ref{EQN005}) from }$\int_{-\pi}^{+\pi}$\textbf{ to }$\int_{-\infty}^{+\infty}$\textbf{:} \\
The relative error is expected to be suppressed exponentially by the second term in the exponential in (\ref{EQN861}) and, therefore, proportional to
\begin{equation}
\exp\bigg(-\frac{\mathcal{E}_2 V}{2} (\pi - \theta_s)^2\bigg) \approx \exp\bigg(-\frac{\pi^2 \mathcal{E}_2 V}{2}\bigg) .
\end{equation}
\end{itemize}
In zero temperature QCD simulations typically $T \gtapprox L$. For sufficiently large values of $m_\pi(\theta) L$, e.g.\
\begin{itemize}
\item[\textbf{(C3)}] $\phantom{xxx} m_\pi(\theta) L \gtapprox 3 \ldots 5 \gg 1$
\end{itemize}
as typically required in QCD simulations, corrections (a) and for the partition function also (b) should essentially be negligible. To be able to ignore corrections (b) for two-point correlation functions, one needs
\begin{itemize}
\item[\textbf{(C4)}] $\phantom{xxx} (M_H^\ast(\theta) - M_H(\theta)) t \gg 1 \quad , \quad M_H(\theta) (T-2 t) \gg 1$.
\end{itemize}
Corrections (c) can be neglected, if $\mathcal{E}_2 V \gg 1$, which is already part of \textbf{(C1)}.

For a discussion of the conditions \textbf{(C1)} to \textbf{(C4)} in the context of a numerical example cf.\ section~\ref{SEC053}.


\section{\label{SEC053}Calculations at fixed topology in quantum mechanics}

To test the equations derived in the previous sections, in particular (\ref{EQN673}) and (\ref{EQN674}), we study a simple model, quantum mechanics on a circle. It can be solved analytically or, in case of a potential, numerically up to arbitrary precision. We extract the difference of the two lowest energy eigenvalues, the equivalent of a hadron mass in QCD, from two-point correlation functions calculated at fixed topology. The insights obtained might be helpful for determining hadron masses from fixed topology simulations in QCD.


\subsection{A particle on a circle in quantum mechanics}

The Lagrangian of a quantum mechanical particle (mass $m$) on a circle (radius $r$) parameterized by the angle $\varphi$ is
\begin{equation}
L \equiv \frac{m r^2}{2} \dot{\varphi}^2 - U(\varphi) = \frac{I}{2} \dot{\varphi}^2 - U(\varphi) ,
\end{equation}
where $I \equiv m r^2$ is the moment of inertia. The potential $U$ will be specified below.

\begin{figure}[htb]
\begin{center}
\includegraphics[scale=1]{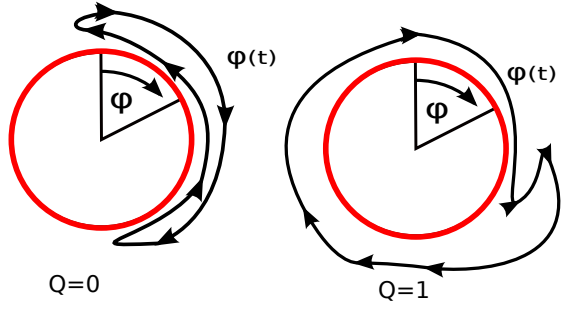} 
\caption{\label{FIG001}paths with topological charge $Q=0$ and $Q=1$.}
\end{center}
\end{figure}

A periodic time with extension $T$ implies $\varphi(t+T) = \varphi(t) + 2 \pi Q$, $Q \in \mathbb{Z}$, and gives rise to topological charge
\begin{equation}
\frac{1}{2 \pi} \int_0^T dt \, \dot{\varphi} = \frac{1}{2 \pi} \Big(\varphi(T) - \varphi(0)\Big) = Q .
\end{equation}
The topological charge density is $q \equiv \dot{\varphi} / 2 \pi$. Exemplary paths with topological charge $Q=0$ and $Q=1$ are sketched in Figure~\ref{FIG001}.

The path integral for the Euclidean partition function is
\begin{equation}
Z \equiv \int D\varphi \, e^{-S_E[\varphi]} \quad , \quad S_E[\varphi] \equiv \int_0^T dt \, L_E \quad , \quad L_E \equiv \frac{I}{2} \dot{\varphi}^2 + U(\varphi) ,
\end{equation}
where the integration $\int D\varphi$ is over all paths, which are $T$-periodic modulo $2 \pi$, i.e.\ over all topological sectors.

The corresponding path integral over a single topological sector $Q$, which is relevant in the context of topology fixing, is
\begin{equation}
\label{EQN672} \begin{aligned}
 & Z_{Q,T} \equiv \int D\varphi \, \delta_{Q,Q(\varphi)} e^{-S_E[\varphi]} = \int D\varphi \, \frac{1}{2 \pi} \int_{-\pi}^{+\pi} d\theta \, e^{i (Q - Q(\varphi)) \theta} e^{-S_E[\varphi]} = \\
 & \hspace{0.7cm} = \frac{1}{2 \pi} \int_{-\pi}^{+\pi} d\theta \, e^{i \theta Q} \underbrace{\int D\varphi \, \exp\bigg(-\underbrace{\bigg(S_E[\varphi] + i \theta \frac{1}{2 \pi} \int_0^T dt \, \dot{\varphi}\bigg)}_{\equiv S_{E,\theta}[\varphi]}\bigg)}_{\equiv \mathcal{Z}_{\theta,T}}
\end{aligned}
\end{equation}
(note that the analog of the spacetime volume $V$ in QCD is in quantum mechanics the temporal extension $T$, i.e.\ throughout this section $V \rightarrow T$). One can read off both $\mathcal{Z}_{\theta,T}$ and $S_{E,\theta}$. The $\theta$-dependent Hamiltonian, which can be obtained as usual, is
\begin{equation}
H_\theta \equiv \frac{1}{2 I} \bigg(p_\varphi + \frac{\theta}{2 \pi}\bigg)^2 + U(\varphi) .
\end{equation}


\subsection{A free particle, $U=0$}


\subsubsection{Eigenfunctions and eigenvalues}

\begin{figure}[htb]
\begin{center}
\includegraphics[scale=1]{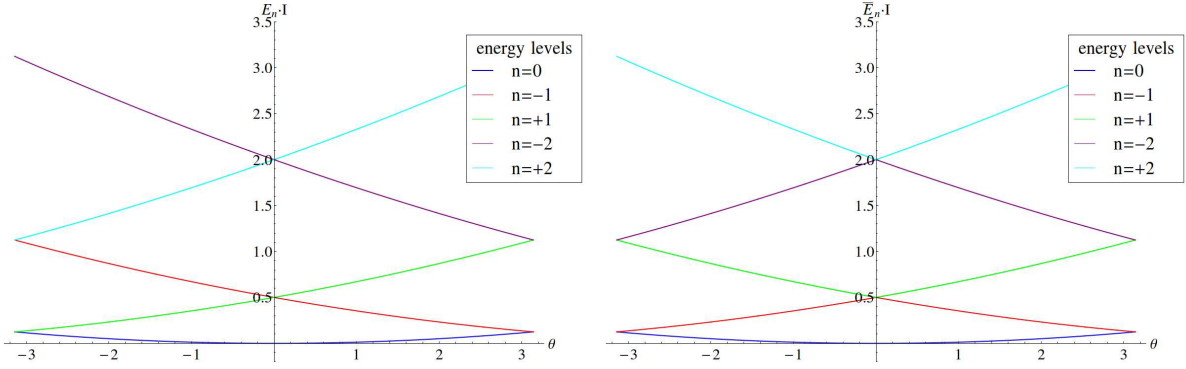}
\caption{\label{FIG022}the low lying spectrum for $U = 0$; \textbf{(left)}~$E_n I$ as a function of $\theta$ (eq.\ (\ref{EQN562})); \textbf{(right)}~$\bar{E}_n I$ as a function of $\theta$ (eq.\ (\ref{EQN563})).}
\end{center}
\end{figure}

For $U = 0$ the eigenfunctions $\psi_n$ and eigenvalues $E_n$ of $H_\theta$ can be determined analytically,
\begin{equation}
\label{EQN562} H_\theta \psi_n(\varphi) = E_n \psi_n(\varphi) \quad \rightarrow \quad \psi_n(\varphi) = \frac{e^{+i n \varphi}}{\sqrt{2 \pi}} \quad , \quad E_n(\theta) = \frac{1}{2 I} \bigg(n + \frac{\theta}{2 \pi}\bigg)^2 .
\end{equation}
Note that in previous sections we used $E_n(+\theta) = E_n(-\theta)$. While the spectrum fulfills this $+\theta \leftrightarrow -\theta$ symmetry, it is clearly violated by our mathematical parameterization (\ref{EQN562}) for $n \neq 0$ (cf.\ Figure~\ref{FIG022}, left plot). An equivalent set of eigenfunctions and eigenvalues fulfilling the $+\theta \leftrightarrow -\theta$ symmetry is
\begin{equation}
\label{EQN563} \bar{\psi}_n(\varphi) = \Theta(+\theta) \frac{e^{+i n \varphi}}{\sqrt{2 \pi}} + \Theta(-\theta) \frac{e^{-i n \varphi}}{\sqrt{2 \pi}} \quad , \quad \bar{E}_n(\theta) = \frac{1}{2 I} \bigg(n + \frac{|\theta|}{2 \pi}\bigg)^2
\end{equation}
(cf.\ Figure~\ref{FIG022}, right plot).

\subsubsection{Partition function}

The partition function $Z_{Q,T}$ is the Fourier transform of $\mathcal{Z}_{\theta,T}$ (cf.\ (\ref{EQN672})). After inserting the eigenvalues $E_n(\theta)$ and changing the variables of integration according to $\theta \rightarrow \theta' = \theta + 2 \pi n$, one obtains a Gaussian integral, which is analytically solvable,
\begin{equation}
\label{EQN798} \begin{aligned}
 & Z_{Q,T} = \frac{1}{2 \pi} \int_{-\pi}^{+\pi} d\theta \, e^{i Q \theta} \mathcal{Z}_{\theta,T} = \frac{1}{2 \pi} \int_{-\pi}^{+\pi} d\theta \, e^{i Q \theta} \sum_n \exp\bigg(- \frac{1}{2 I} \bigg(n + \frac{\theta}{2 \pi}\bigg)^2 T\bigg) \ \ = \\
 & \hspace{0.7cm} = \frac{1}{2 \pi} \underbrace{\sum_n \int_{-\pi - 2 \pi n}^{+\pi - 2 \pi n} d\theta'}_{= \int_{-\infty}^{+\infty} d\theta'} \, e^{i Q \theta'} \exp\bigg(-\frac{T}{8 \pi^2 I} \theta'^2\bigg) = \sqrt{\frac{2 \pi I}{T}} \exp\bigg(-\frac{2 \pi^2 I}{T} Q^2\bigg) .
\end{aligned}
\end{equation}

This exact result can be compared with the approximation (\ref{EQN010}), after inserting $E_0(0,V_s) \rightarrow E_0(\theta=0) = \theta^2 / 8 \pi^2 I|_{\theta = 0} = 0$, $\mathcal{E}_2 = E_0^{(2)}(\theta=0) = 1 / 4 \pi^2 I$ and $\mathcal{E}_n = 0$ for $n \neq 2$,
\begin{equation}
\label{EQN628} Z_{Q,T} = \sqrt{\frac{2 \pi I}{T}} \exp\bigg(-\frac{2 \pi^2 I}{T} Q^2\bigg) + \mathcal{O}\bigg(\frac{1}{\mathcal{E}_2^4 T^4} \ , \ \frac{1}{\mathcal{E}_2^4 T^4} Q^2 \ , \ \frac{1}{\mathcal{E}_2^4 T^4} Q^4\bigg) .
\end{equation}
Even though power corrections proportional to $1/T^4$ and exponentially suppressed corrections have been neglected, the approximation is identical to the exact result (\ref{EQN798}).


\subsubsection{Two-point correlation function}

We use the creation operator $O \equiv \sin(\varphi)$ (on a circle operators must be $2 \pi$-periodic in $\varphi$). Note that
\begin{itemize}
\item $O | \bar{\psi}_0 ; \theta \rangle$ is orthogonal to the ground state $| \bar{\psi}_0 ; \theta \rangle$, which is required for (\ref{EQN856}) (and consequently for (\ref{EQN673}) and (\ref{EQN674})) to be valid,

\item $O | \bar{\psi}_0 ; \theta \rangle$ has non-vanishing overlap to the first excitation $| \bar{\psi}_{-1} ; \theta \rangle$,
\end{itemize}
i.e.\ $O$ is a suitable creation operator for the first excitation $| \bar{\psi}_{-1} ; \theta \rangle$.

The two-point correlation function $C_{Q,T}(t)$ is the Fourier transform of $\mathcal{C}_{\theta,T}(t)$, which can be expanded in terms of energy eigenstates (cf.\ (\ref{EQN415}) and (\ref{EQN597})). After inserting the eigenvalues $E_n(\theta)$ (eq.\ (\ref{EQN562})), using $\langle \psi_m ; \theta | O | \psi_n ; \theta \rangle = (\delta_{m,n+1} + \delta_{m,n-1}) / 2$ and changing the variables of integration according to $\theta \rightarrow \theta + 2 \pi n$ as in (\ref{EQN798}), one again obtains a Gaussian integral, which can be solved exactly,
\begin{equation}
\begin{aligned}
\label{EQN958} C_{Q,T}(t) = \frac{1}{2} \exp\bigg(-\frac{t (T-t)}{2 I T}\bigg) \cos\bigg(\frac{2 \pi Q t}{T}\bigg) .
\end{aligned}
\end{equation}

The analog of the lightest hadron mass in QCD is the difference of the energy eigenvalues of the first excitation and the ground state,
\begin{equation}
M_H(\theta) = \bar{E}_{-1}(\theta) - \bar{E}_0(\theta) = \frac{1 - |\theta| / 2\pi}{2 I} .
\end{equation}
Clearly $M_H^{(2)}(0) = \infty$, which implies $x_2 = \infty$. This in turn severely violates condition \textbf{(C2)} of section~\ref{EQN735}, which was assumed to be fulfilled, when deriving the approximations of two-point correlation functions (\ref{EQN673}) and (\ref{EQN674}). In other words, agreement between the exact result (\ref{EQN958}) and (\ref{EQN673}) and (\ref{EQN674}) cannot be expected and is neither observed.

To circumvent the problem, one can use the eigenvalue parameterization (\ref{EQN562}), which, however, does not fulfill $E_n(+\theta) = E_n(-\theta)$. The consequence is that the expansion (\ref{EQN079}) may also contain odd terms $\mathcal{F}_1 \theta$, $\mathcal{F}_3 \theta^3 / 6$, etc. For a free particle, however, only few parameters are non-zero,
\begin{itemize}
\item $E_0(\theta) = \theta^2 / 8 \pi^2 I \\ \rightarrow \quad \mathcal{E}_2 = 1 / 4 \pi^2 I$,

\item $M_{H,\pm 1}(\theta) \equiv E_{\pm 1}(\theta)- E_0(\theta) = (1 \pm \theta / \pi) / 2 I$ \\ (the two lightest hadron masses need to be considered, since $M_{H,+1}(\theta) < M_{H,-1}(\theta)$ for $\theta < 0$ and $M_{H,+1}(\theta) > M_{H,-1}(\theta)$ for $\theta > 0$; cf.\ Figure~\ref{FIG022}) \\ $\rightarrow \quad M_{H,\pm 1}(0) = 1 / 2 I \ , \ M_{H,\pm 1}^{(1)}(0) = \pm 1 / 2 \pi I$,

\item $\alpha(\theta) = 1/4 \\ \rightarrow \quad \alpha(0) = 1/4$.
\end{itemize}
All further parameters $\mathcal{E}_n$, $M_{H,\pm 1}^{(n)}(0)$ and $\beta^{(n)}$ vanish. Consequently, $\mathcal{F}_0 = M_{H,\pm 1}(0) t / T$, $\mathcal{F}_1 = M_{H,\pm 1}^{(1)}(0) t / T$ and $\mathcal{F}_2 = \mathcal{E}_2$, while $\mathcal{F}_n = 0$ for $n \geq 3$. In other words in (\ref{EQN079}) there is only a single additional term, $\mathcal{F}_1 \theta$. Since this term is proportional to $\theta$, and since there is already a term proportional to $\theta$ in (\ref{EQN079}), $-i Q \theta$, it can easily be included in the calculation from section~\ref{SEC459} by replacing $Q \rightarrow Q + i M_{H,+1}^{(1)}(0) t$ and $Q \rightarrow Q + i M_{H,-1}^{(1)}(0) t$ in (\ref{EQN624}), respectively, and by adding both results, to obtain $C_{Q,V}(t) Z_{Q,V}$. Inserting the above listed parameters and dividing by $Z_{Q,V}$ (eq.\ (\ref{EQN628})) one finds
\begin{equation}
\begin{aligned}
C_{Q,T}(t) = \frac{1}{2} \exp\bigg(-\frac{t (T-t)}{2 I T}\bigg) \cos\bigg(\frac{2 \pi Q t}{T}\bigg) + \mathcal{O}\bigg(\frac{1}{\mathcal{E}_2^4 T^4} \ , \ \frac{1}{\mathcal{E}_2^4 T^4} Q^2 \ , \ \frac{1}{\mathcal{E}_2^4 T^4} Q^4\bigg) ,
\end{aligned}
\end{equation}
which is identical to the exact result (\ref{EQN958}), even though power corrections proportional to $1/T^4$ and exponentially suppressed corrections have been neglected.

The problems associated with $M_{H,\pm}^{(2)}(0) = \infty$, do not appear, when a potential $U \neq 0$ is chosen (cf.\ section~\ref{SEC694} and Figure~\ref{FIG023}). They are also not expected to be present in QCD.


\subsection{\label{SEC694}A particle in a square well}

Now we study a square well potential
\begin{equation}
\label{EQN427} U(\varphi) \equiv \begin{cases} 0 & \textrm{if }-\rho/2 < \varphi < +\rho/2 \\ U_0 & \textrm{otherwise} \end{cases}
\end{equation}
($U_0 > 0$ is the depth and $\rho > 0$ the width of the well). Again we use the creation operator $O \equiv \sin(\varphi)$, for which one can show $\langle 0 ; \theta | O | 0 ; \theta \rangle = 0$ \footnote{A complicated theory like QCD has many symmetries and, therefore, many orthogonal sectors of states, which are labeled by the corresponding quantum numbers (total angular momentum, charge conjugation, flavor quantum numbers). In such a theory one typically chooses an operator exciting states, which do not have the quantum numbers of the vacuum, i.e.\ where $\langle 0 ; \theta , V_s | O | 0 ; \theta , V_s \rangle = 0$, due to symmetry. In the simple quantum mechanical model parity is the only symmetry, which is broken at $\theta \neq 0$. Therefore, constructing an appropriate creation operator is less straightforward, because $\langle 0 ; \theta | O | 0 ; \theta \rangle = 0$ is not guaranteed by obvious symmetries, but has to be shown explicitly.} (cf.\ appendix~\ref{SEC635}).


\subsubsection{Solving the model numerically}

For the square well potential (\ref{EQN427}) the Schr\"odinger equation cannot be solved analytically, but numerically up to arbitrary precision, i.e.\ no simulations are required. For these numerical computations we express all dimensionful quantities in units of $I$, i.e.\ we work with dimensionless quantities (denoted by a hat $\hat{\phantom{x}}$) $I \rightarrow \hat{I} = I / I = 1$, $T \rightarrow \hat{T} = T / I$ and $U_0 \rightarrow \hat{U}_0 = U_0 I$. For the numerical results presented in this section we have used $\hat{U}_0 = 5.0$ and $\rho = 0.9 \times 2 \pi$.

\begin{figure}[htb]
\begin{center}
\includegraphics[scale=1]{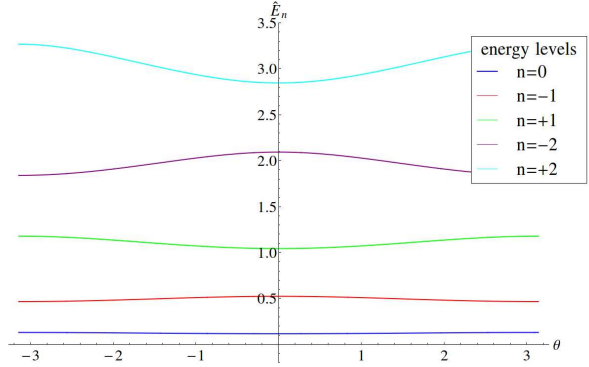}
\caption{\label{FIG023}the low lying energy eigenvalues $\hat{E}_n$ for the square well potential (\ref{EQN427}) with $\hat{U}_0 = 5.0$ and $\rho = 0.9 \times 2 \pi$ as functions of $\theta$.}
\end{center}
\end{figure}

We proceeded as follows:
\begin{enumerate}
\item Solve Schr\"odinger's equation
\begin{eqnarray}
\hat{H}_\theta \psi_n(\varphi;\theta) = \hat{E}_n(\theta) \psi_n(\varphi;\theta) \quad , \quad \hat{H}_\theta = \frac{1}{2} \bigg(p_\varphi + \frac{\theta}{2 \pi}\bigg)^2 + \hat{U}(\varphi)
\end{eqnarray}
($\hat{H}_\theta \equiv H_\theta I$, $\hat{E}_n(\theta) = E_n(\theta) I$, $\hat{U} \equiv U I$) as outlined in appendix~\ref{SEC634}. The resulting low lying spectrum is shown in Figure~\ref{FIG023}.

\item Use the resulting energy eigenvalues $\hat{E}_0(\theta)$ and $\hat{E}_1(\theta)$ to determine
\begin{itemize}
\item $\hat{\mathcal{E}}_n = \hat{E}_0^{(n)}(0)$, $n = 0, 2, 4, 6, 8$.

\item $\hat{M}_H^{(n)}(0) = (d / d\theta)^n (\hat{E}_1(\theta)-\hat{E}_0(\theta))|_{\theta = 0}$, $n = 0, 2, 4, 6, 8$
\end{itemize}
and the resulting wave functions $\psi_0(\varphi;\theta)$ and $\psi_1(\varphi;\theta)$ to determine
\begin{itemize}
\item $\alpha^{(n)}(0)$, $n = 0, 2, 4, 6, 8$,

\item $\beta^{(n)}(0)$, $n = 2, 4, 6, 8$,
\end{itemize}
where
\begin{equation}
\alpha(\theta) = \bigg|\int_0^{2 \pi} d\varphi \, (\psi_1(\varphi;\theta))^\ast \sin(\varphi) \psi_0(\varphi;\theta)\bigg|^2 \quad , \quad \beta(\theta) = -\ln\bigg(\frac{\alpha(\theta)}{\alpha(0)}\bigg) .
\end{equation}
These are the parameters of the two-point correlation function $C_{Q,\hat{T}}(\hat{t})$, $\hat{t} = t/I$ (eqs.\ (\ref{EQN673}) and (\ref{EQN674})). For $\hat{U}_0 = 5.0$ and $\rho = 0.9 \times 2 \pi$ they are collected in Table~\ref{TAB001}.

\item Calculate $\mathcal{C}_{\theta,T}(\hat{t})$ using sufficiently many low lying energy eigenvalues and corresponding wave functions from step~1 such that the exponentially suppressed error is negligible already for very small temporal separations (cf.\ eqs.\ (\ref{EQN001}) and (\ref{EQN597})).

\item Perform a Fourier transformation numerically to obtain $C_{Q,\hat{T}}(\hat{t})$, the exact correlation function at fixed topology.

\item Define and calculate the effective mass
\begin{equation}
\label{eq:M_ex} \hat{M}_{Q,\hat{T}}^\textrm{eff}(\hat{t}) \equiv -\frac{d}{d\hat{t}} \ln\Big(C_{Q,\hat{T}}(\hat{t})\Big) .
\end{equation}
\end{enumerate}

\begin{table}[htb]
\begin{center}
\begin{tabular}{|c||c|c|c|c|}
\hline
 & & & & \vspace{-0.40cm} \\
$n$ & $\hat{\mathcal{E}}_n$ & $\hat{M}_H^{(n)}(0)$ & $\alpha^{(n)}(0)$ & $\beta^{(n)}(0)$ \\
 & & & & \vspace{-0.40cm} \\
\hline
 & & & & \vspace{-0.40cm} \\
$0$ & $+0.11708$ & $+0.40714$ & $+0.50419$ &  \\
$2$ & $+0.00645$ & $-0.03838$ & $-0.00357$ & $+0.00709$ \\
$4$ & $-0.00497$ & $+0.04983$ & $+0.00328$ & $-0.00636$ \\
$6$ & $+0.00042$ & $-0.13191$ & $-0.04721$ & $+0.09308$ \\
$8$ & $+0.00834$ & $+0.95631$ & $+0.91037$ & $-1.77931$\vspace{-0.40cm} \\
 & & & & \\
\hline
\end{tabular}

\caption{\label{TAB001}the parameters of the two-point correlation function $C_{Q,\hat{T}}(\hat{t})$ (eqs.\ (\ref{EQN673}) and (\ref{EQN674})) for $\hat{U}_0 = 5.0$ and $\rho = 0.9 \times 2 \pi$.}

\end{center}
\end{table}


\subsubsection{Effective masses at fixed topology}

\begin{figure}[htb]
\begin{center}
\includegraphics[scale=1]{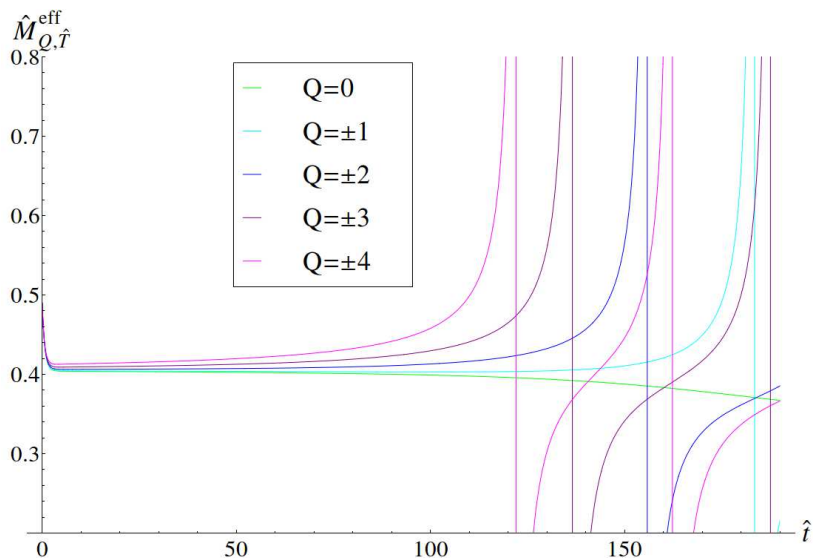}
\caption{\label{fig:M(t)}effective masses $\hat{M}_{Q,\hat{T}}^\textrm{eff}$ as functions of the temporal separation $\hat{t}$ for different topological sectors $Q$ and $\hat{T} = 6.0 / \hat{\mathcal{E}}_2 \approx 930.2$.}
\end{center}
\end{figure}

In Figure~\ref{fig:M(t)} we show effective masses $\hat{M}_{Q,\hat{T}}^\textrm{eff}$ (eq.\ (\ref{eq:M_ex})) as functions of the temporal separation $\hat{t}$ for different topological sectors $Q$ and $\hat{T} = 6.0 / \hat{\mathcal{E}}_2 \approx 930.2$. As usual at small temporal separations the effective masses are quite large and strongly decreasing, due to the presence of excited states. At large temporal separations there are also severe deviations from a constant behavior. This contrasts ordinary quantum mechanics or quantum field theory (i.e.\ at unfixed topology) and is caused by topology fixing. This effect is also visible in the $1/V$ expansion of the two point correlation function, in particular in (\ref{EQN674}), where the exponent is not purely linear in $t$ for large $t$, but contains also terms proportional to $t^2$ and $t^3$. At intermediate temporal separations there are plateau-like regions, which become smaller with increasing topological charge $Q$.

\subsubsection{\label{SEC184}Comparison of the $1/V$ expansions of $C_{Q,T}(t)$ and the exact result}

\begin{figure}[h!]
\begin{center}
\includegraphics[scale=1]{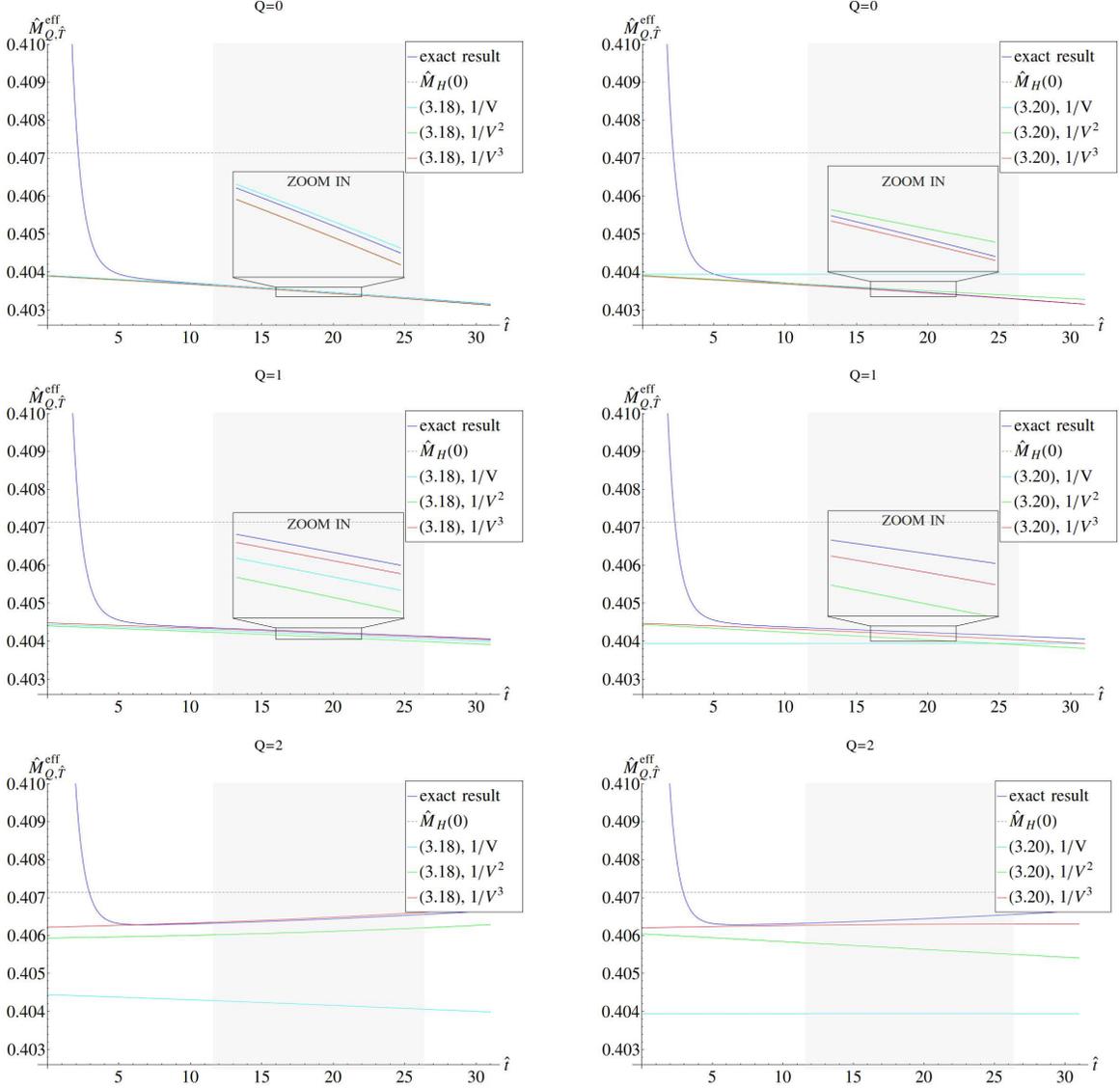}
\caption{\label{FIG099}effective masses $\hat{M}_{Q,\hat{T}}^\textrm{eff}$ derived from the $1/V$ expansions of two-point correlation functions as functions of the temporal separation $\hat{t}$ for different topological sectors $Q$ and $\hat{T} = 6.0 / \hat{\mathcal{E}}_2 \approx 930.2$.}
\end{center}
\end{figure}

In Figure~\ref{FIG099} we show effective masses derived from the $1/V$ expansions of two-point correlation functions\footnote{Note that in quantum mechanics a $1/V$ expansion is a $1/T$ expansion.} (\ref{EQN673}) (left column) and (\ref{EQN674}) (right column) using the definition (\ref{eq:M_ex}). The first, second and third row correspond to $Q=0$, $|Q| = 1$ and $|Q| = 2$, respectively. To illustrate the relative importance of $1/V$, $1/V^2$ and $1/V^3$ terms, we also show versions of (\ref{EQN673}) and (\ref{EQN674}), which are only derived up to $\mathcal{O}(1/V)$ and $\mathcal{O}(1/V^2)$. While less accurate, these expressions contain a smaller number of parameters, which might be an advantage, when e.g.\ fitting to results from lattice simulations (such a fitting is discussed in section~\ref{SEC439}). In detail the following curves are shown with $V \rightarrow T$ and the parameters taken from Table~\ref{TAB001}:
\begin{itemize}
\item $\hat{M}_{Q,\hat{T}}^\textrm{eff}(\hat{t})$ from (\ref{EQN673}), derived up to $\mathcal{O}(1/V)$:
\begin{equation}
\label{EQN673V1} \begin{aligned}
 & C_{Q,V}(t) = \frac{\alpha(0)}{\sqrt{1 + x_2 / \mathcal{E}_2 V}} \exp\bigg(-M_H(0) t - \frac{1}{\mathcal{E}_2 V} \bigg(\frac{1}{1 + x_2 / \mathcal{E}_2 V} - 1\bigg) \frac{1}{2} Q^2\bigg) \frac{G_\mathcal{C}}{G} \\
 & G_\mathcal{C} = 1 - \frac{1}{\mathcal{E}_2 V} \frac{\mathcal{E}_4 (1 + x_4 / \mathcal{E}_4 V)}{8 \mathcal{E}_2 (1 + x_2 / \mathcal{E}_2 V)^2} \\
 & G = 1 - \frac{1}{\mathcal{E}_2 V} \frac{\mathcal{E}_4}{8 \mathcal{E}_2} ;
\end{aligned}
\end{equation}
\textbf{8 parameters} ($\mathcal{E}_2$, $\mathcal{E}_4$, $M_H(0)$, $M_H^{(2)}(0)$, $M_H^{(4)}(0)$, $\alpha(0)$, $\beta^{(2)}(0)$, $\beta^{(4)}(0)$).

\item $\hat{M}_{Q,\hat{T}}^\textrm{eff}(\hat{t})$ from (\ref{EQN673}), derived up to $\mathcal{O}(1/V^2)$:
\begin{equation}
\label{EQN673V2} \begin{aligned}
 & C_{Q,V}(t) = \frac{\alpha(0)}{\sqrt{1 + x_2 / \mathcal{E}_2 V}} \\
 & \hspace{1.4cm} \exp\bigg(-M_H(0) t - \frac{1}{\mathcal{E}_2 V} \bigg(\frac{1}{1 + x_2 / \mathcal{E}_2 V} - 1\bigg) \frac{1}{2} Q^2\bigg) \\
 & \hspace{1.4cm} \bigg(1 - \frac{1}{(\mathcal{E}_2 V)^2} \frac{\mathcal{E}_4}{2 \mathcal{E}_2} Q^2\bigg)^{+1/2} \bigg(1 - \frac{1}{(\mathcal{E}_2 V)^2} \frac{\mathcal{E}_4 (1 + x_4 / \mathcal{E}_4 V)}{2 \mathcal{E}_2 (1 + x_2 / \mathcal{E}_2 V)^3} Q^2\bigg)^{-1/2} \frac{G_\mathcal{C}}{G} \\
 & G_\mathcal{C} = 1 - \frac{1}{\mathcal{E}_2 V} \frac{\mathcal{E}_4 (1 + x_4 / \mathcal{E}_4 V)}{8 \mathcal{E}_2 (1 + x_2 / \mathcal{E}_2 V)^2} \\
 & \hspace{0.7cm} + \frac{1}{(\mathcal{E}_2 V)^2} \bigg(-\frac{\mathcal{E}_6 (1 + x_6 / \mathcal{E}_6 V)}{48 \mathcal{E}_2 (1 + x_2 / \mathcal{E}_2 V)^3} + \frac{35 \mathcal{E}_4^2 (1 + x_4 / \mathcal{E}_4 V)^2}{384 \mathcal{E}_2^2 (1 + x_2 / \mathcal{E}_2 V)^4}\bigg) \\
 & G = 1 - \frac{1}{\mathcal{E}_2 V} \frac{\mathcal{E}_4}{8 \mathcal{E}_2} + \frac{1}{(\mathcal{E}_2 V)^2} \bigg(-\frac{\mathcal{E}_6}{48 \mathcal{E}_2} + \frac{35 \mathcal{E}_4^2}{384 \mathcal{E}_2^2}\bigg) ;
\end{aligned}
\end{equation}
\textbf{11 parameters} ($\mathcal{E}_2$, $\mathcal{E}_4$, $\mathcal{E}_6$, $M_H(0)$, $M_H^{(2)}(0)$, $M_H^{(4)}(0)$, $M_H^{(6)}(0)$, $\alpha(0)$, $\beta^{(2)}(0)$, $\beta^{(4)}(0)$, $\beta^{(6)}(0)$).

\item $\hat{M}_{Q,\hat{T}}^\textrm{eff}(\hat{t})$ from (\ref{EQN673}) (which is derived up to $\mathcal{O}(1/V^3)$); \\
\textbf{14 parameters} ($\mathcal{E}_2$, $\mathcal{E}_4$, $\mathcal{E}_6$, $\mathcal{E}_8$, $M_H(0)$, $M_H^{(2)}(0)$, $M_H^{(4)}(0)$, $M_H^{(6)}(0)$, $M_H^{(8)}(0)$, $\alpha(0)$, $\beta^{(2)}(0)$, $\beta^{(4)}(0)$, $\beta^{(6)}(0)$, $\beta^{(8)}(0)$).

\item $\hat{M}_{Q,\hat{T}}^\textrm{eff}(\hat{t})$ from (\ref{EQN674}), up to $\mathcal{O}(1/V)$:
\begin{equation}
\label{EQN674V1} C_{Q,V}(t) = \alpha(0) \exp\bigg(-M_H(0) t - \frac{1}{\mathcal{E}_2 V} \frac{x_2}{2}\bigg) ;
\end{equation}
\textbf{5 parameters} ($\mathcal{E}_2$, $M_H(0)$, $M_H^{(2)}(0)$, $\alpha(0)$, $\beta^{(2)}(0)$).

\item $\hat{M}_{Q,\hat{T}}^\textrm{eff}(\hat{t})$ from (\ref{EQN674}), up to $\mathcal{O}(1/V^2)$:
\begin{equation}
\label{EQN674V2} \begin{aligned}
 & C_{Q,V}(t) = \\
 & \hspace{0.7cm} = \alpha(0) \exp\bigg(-M_H(0) t - \frac{1}{\mathcal{E}_2 V} \frac{x_2}{2} - \frac{1}{(\mathcal{E}_2 V)^2} \bigg(\frac{x_4 - 2 (\mathcal{E}_4/\mathcal{E}_2) x_2 - 2 x_2^2}{8} - \frac{x_2}{2} Q^2\bigg)\bigg) ;
\end{aligned}
\end{equation}
\textbf{8 parameters} ($\mathcal{E}_2$, $\mathcal{E}_4$, $M_H(0)$, $M_H^{(2)}(0)$, $M_H^{(4)}(0)$, $\alpha(0)$, $\beta^{(2)}(0)$, $\beta^{(4)}(0)$).

\item $\hat{M}_{Q,\hat{T}}^\textrm{eff}(\hat{t})$ from (\ref{EQN674}) (which is derived up to $\mathcal{O}(1/V^3)$); \\
\textbf{11 parameters} ($\mathcal{E}_2$, $\mathcal{E}_4$, $\mathcal{E}_6$, $M_H(0)$, $M_H^{(2)}(0)$, $M_H^{(4)}(0)$, $M_H^{(6)}(0)$, $\alpha(0)$, $\beta^{(2)}(0)$, $\beta^{(4)}(0)$, $\beta^{(6)}(0)$).
\end{itemize}
Note that the definition (\ref{eq:M_ex}) of $\hat{M}_{Q,\hat{T}}^\textrm{eff}(\hat{t})$ eliminates $\alpha(0)$, i.e.\ effective masses have one parameter less than the corresponding two-point correlation functions. For comparison we also include the exact result already shown and discussed in Figure~\ref{fig:M(t)}. Finally, the dashed line indicates the ``hadron mass'' $\hat{M}_H(0)$ at unfixed topology, to demonstrate the effect of topology fixing on effective masses.

The validity of the shown $1/V$ expansions has been discussed in section~\ref{SEC867} and summarized in terms of four conditions, which we check for the quantum mechanical example with parameters $\hat{U}_0 = 5.0$, $\rho = 0.9 \times 2 \pi$ and $\hat{T} = 6.0 / \hat{\mathcal{E}}_2$:
\begin{itemize}
\item \textbf{(C1)}: \\
$1 \gg 1/\mathcal{E}_2 T = 1/6.0$ and $1 \gg |Q|/\mathcal{E}_2 T = 0 \ , \ 1/6.0 \ , \ 1/3.0$ for $|Q| = 0 \ , \ 1 \ , \ 2$, i.e.\ fulfilled. $|Q| = 3, 4, \ldots$ might need a larger $T$ extension.

\item \textbf{(C2)}: \\
Solving \textbf{(C2)}, $|x_2| = |M_H^{(2)}(0) t + \beta^{(2)}(0)| \ltapprox 1$, with respect to $t$ and inserting the numbers from Table~\ref{TAB001} yields $\hat{t} \ltapprox |(1 + \beta^{(2)}(0)) / \hat{M}_H^{(2)}(0)| \approx 26.2$. For significantly larger $\hat{t}$ values the accuracy of the $1/V$ expansions is expected to suffer. The ``safe region'' $\hat{t} \ltapprox 26.2$ is shaded in light gray in Figure~\ref{FIG099}.

\item \textbf{(C3)}: \\
Figure~\ref{FIG023} shows that $\hat{M}_H(\theta)$ (the analog of $m_\pi$ in QCD) is minimal at $\theta = \pm \pi$, $\hat{M}_H(\pm \pi) = 0.336$. $m_\pi L$ corresponds to $\hat{M}_H(\pm \pi) \hat{T}$ and $\hat{M}_H(\pm \pi) \hat{T} = 6.0 \times \hat{M}_H(\pm \pi) / \hat{\mathcal{E}}_2 \approx 312.6 \gg 1$, i.e.\ the condition is clearly fulfilled.

\item \textbf{(C4)}: \\
Figure~\ref{FIG023} shows that $\hat{M}_H^\ast(\theta) - \hat{M}_H(\theta)$ is minimal at $\theta = 0$, $\hat{M}_H^\ast(0) - \hat{M}_H(0) = 0.520$; therefore, $(M_H^\ast(0) - M_H(0)) t \gg 1$ corresponds to $\hat{t} \gg 1 / (\hat{M}_H^\ast(0) - \hat{M}_H(0)) \approx 1.92$. We consider $6.0 \gg 1$ and shade the corresponding safe region $\hat{t} > 6.0 \times 1.920 \approx 11.5$ in light gray. \\
Finally $M_H(\theta) (T-2t) \gg 1$ can be solved with respect to $\hat{t}$ resulting in $\hat{t} \ll (\hat{T} - 1 / \hat{M}_H(\theta)) / 2 \approx 463.6$. Clearly also this condition is fulfilled.
\end{itemize}
The effective mass plots shown in Figure~\ref{FIG099} are consistent with these estimates. There is nearly perfect agreement between the $1/V$ expansions of $\hat{M}_{Q,\hat{T}}^\textrm{eff}(\hat{t})$ and the exact results in the gray regions. On the other hand the difference of the effective mass at fixed topology and the mass at unfixed topology (the quantity one is finally interested in) is quite large. This clearly indicates that determining hadron masses from fixed topology simulations with standard methods (e.g.\ fitting a constant to an effective mass at large temporal separations) might lead to sizable systematic errors, which, however, can be reduced by orders of magnitude, when using the discussed $1/V$ expansions of $\hat{M}_{Q,\hat{T}}^\textrm{eff}(\hat{t})$.

\begin{figure}[htb]
\begin{center}
\includegraphics[scale=1]{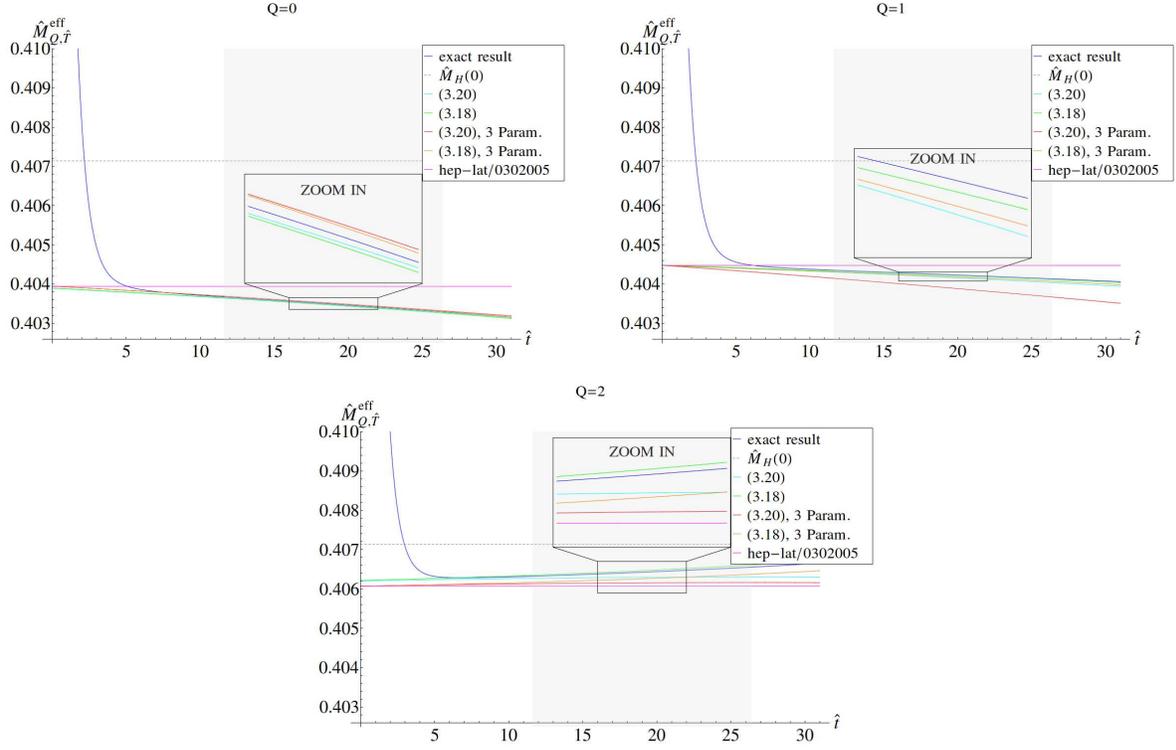}
\caption{\label{FIG199}effective masses $\hat{M}_{Q,\hat{T}}^\textrm{eff}$ derived from (\ref{EQN673}) and (\ref{EQN674}) restricted to the parameters $\mathcal{E}_2$, $M_H(0)$ and $M_H^{(2)}(0)$ as functions of the temporal separation $\hat{t}$ for different topological sectors $Q$ and $\hat{T} = 6.0 / \hat{\mathcal{E}}_2 \approx 930.2$.}
\end{center}
\end{figure}

The number of parameters, in particular for the expansions derived up to $\mathcal{O}(1/V^3)$, i.e.\ (\ref{EQN673}) and (\ref{EQN674}), is quite large. This could be a problem, when fitting these expressions to lattice results for two-point correlation functions, where statistical accuracy is limited, e.g.\ for expensive QCD simulations. A possibility to benefit from the higher order expansions at least to some extent, while keeping at the same time the number of fit parameters small, is to use eqs.\ (\ref{EQN673}) and (\ref{EQN674}) (i.e.\ expansions up to $\mathcal{O}(1/V^3)$), but to set parameters, which are expected to be less important, to zero. In Figure~\ref{FIG199} we explore this possibility by restricting (\ref{EQN673}) and (\ref{EQN674}) to the parameters $\mathcal{E}_2$, $M_H(0)$, $M_H^{(2)}(0)$ and $\alpha(0)$, which are the \textbf{4 parameters} of eq.\ (\ref{EQN708}), the $1/V$ expansion from the seminal paper \cite{Brower:2003yx}. In detail the following curves are shown with the parameters taken from Table~\ref{TAB001}:
\begin{itemize}
\item $\hat{M}_{Q,\hat{T}}^\textrm{eff}(\hat{t})$ from (\ref{EQN673}).

\item $\hat{M}_{Q,\hat{T}}^\textrm{eff}(\hat{t})$ from (\ref{EQN673}), restricted to the \textbf{3 parameters} $\mathcal{E}_2$, $M_H(0)$ and $M_H^{(2)}(0)$:
\begin{equation}
\label{EQN900} C_{Q,V}(t) = \frac{\alpha(0)}{\sqrt{1 + x_2 / \mathcal{E}_2 V}} \exp\bigg(-M_H(0) t - \frac{1}{\mathcal{E}_2 V} \bigg(\frac{1}{1 + x_2 / \mathcal{E}_2 V} - 1\bigg) \frac{1}{2} Q^2\bigg)
\end{equation}
with $x_2 \equiv M_H^{(2)} t$.

\item $\hat{M}_{Q,\hat{T}}^\textrm{eff}(\hat{t})$ from (\ref{EQN674}).

\item $\hat{M}_{Q,\hat{T}}^\textrm{eff}(\hat{t})$ from (\ref{EQN674}), restricted to the \textbf{3 parameters} $\mathcal{E}_2$, $M_H(0)$ and $M_H^{(2)}(0)$:
\begin{equation}
\label{EQN900_} \begin{aligned}
 & C_{Q,V}(t) = \alpha(0) \\
 & \hspace{1.4cm} \exp\bigg(-M_H(0) t - \frac{1}{\mathcal{E}_2 V} \frac{x_2}{2} + \frac{1}{(\mathcal{E}_2 V)^2} \bigg(\frac{x_2^2}{4} + \frac{x_2}{2} Q^2\bigg) - \frac{1}{(\mathcal{E}_2 V)^3} \bigg(\frac{x_2^3}{6} + \frac{x_2^2}{2} Q^2\bigg)\bigg)
\end{aligned}
\end{equation}
with $x_2 \equiv M_H^{(2)} t$.

\item $\hat{M}_{Q,\hat{T}}^\textrm{eff}(\hat{t})$ from (\ref{EQN708}), the $1/V$ expansion from \cite{Brower:2003yx}.
\end{itemize}
Even though the number of parameters is identical, the ``parameter restricted $\mathcal{O}(1/V^3)$ expansions'', in particular (\ref{EQN900}), are significantly closer to the exact result. In practice, when fitting to a correlator from fixed topology QCD simulations with statistical errors, where one is limited in the number of fit parameters, using (\ref{EQN900}) might be the best compromise.

\subsubsection{\label{SEC439}Extracting hadron masses from fixed topology simulations}

A straightforward method to determine physical hadron masses (i.e.\ hadron masses at unfixed topology) from fixed topology simulations based on the $1/V$ expansion (\ref{EQN708}) and (\ref{EQN709}) has been proposed in \cite{Brower:2003yx}:
\begin{enumerate}
\item Perform simulations at fixed topology for different topological charges $Q$ and spacetime volumes $V$. Determine ``fixed topology hadron masses'' $M_{Q,V}$ (denoted by $M_Q$ in (\ref{EQN708}) and \cite{Brower:2003yx}) using (\ref{EQN708}) for each simulation.

\item Determine the hadron mass $M_H(0)$ (the hadron mass at unfixed topology), $M_H^{(2)}(0)$ and $\mathcal{E}_2 = \chi_t$ by fitting (\ref{EQN709}) to the fixed topology hadron masses $M_{Q,V}$ obtained in step~1.
\end{enumerate}

Note, however, that two-point correlation functions at fixed topology do not decay exponentially $\propto e^{-M_{Q,V} t}$ at large temporal separations $t$ (cf.\ e.g.\ (\ref{EQN673})), as their counterparts at unfixed topology do. Therefore, determining a fixed topology and finite volume mass $M_{Q,V}$ is not clear without ambiguity. One could e.g.\ define $M_{Q,V}$ at some temporal separation $t_M$, where the $1/V$ expansion is a good approximation, i.e.\ where the conditions \textbf{(C2)} and \textbf{(C4)} from section~\ref{SEC867} are fulfilled, using (\ref{eq:M_ex}), i.e.\
\begin{equation}
\label{EQN589} M_{Q,V} \equiv M_{Q,V}^\textrm{eff}(t_M) = -\frac{d}{dt} \ln\Big(C_{Q,V}(t)\Big)\Big|_{t=t_M} .
\end{equation}

\begin{figure}[htb]
\begin{center}
\includegraphics[scale=1]{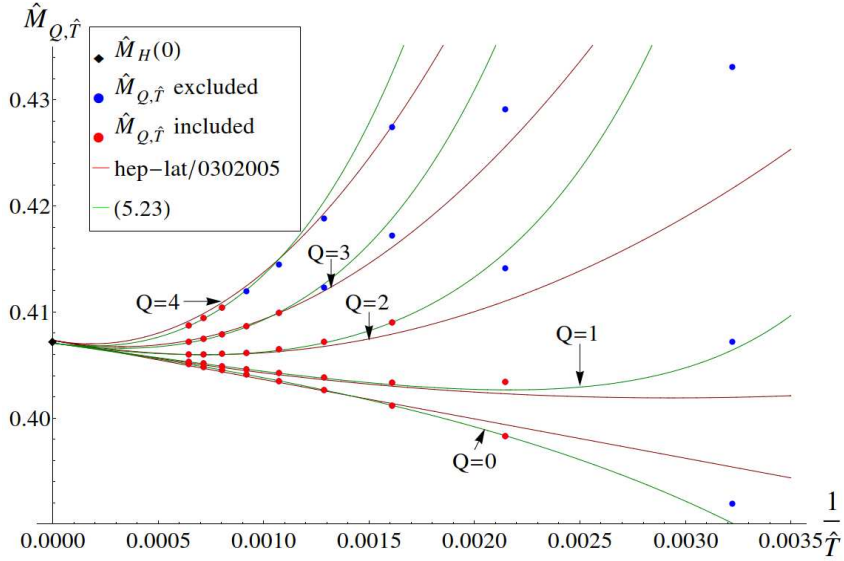}
\caption{\label{FIG340}determining the physical mass $\hat{M}_H(0)$ (i.e.\ the mass at unfixed topology) from a fixed topology computation; only those masses $\hat{M}_{Q,\hat{T}}$ are included in the fit, which fulfill $1 / \hat{\mathcal{E}}_2 T , |Q| / \hat{\mathcal{E}}_2 T \leq 0.5$ (red points).}
\end{center}
\end{figure}

We now follow this strategy to mimic the method to determine a physical hadron mass (i.e.\ at unfixed topology) from fixed topology computations using the quantum mechanical model. To this end we choose $\hat{t}_M = 20.0$, i.e.\ a $\hat{t}_M$ value inside the ``safe gray regions'' of Figure~\ref{FIG099} and Figure~\ref{FIG199}. We use the exact result for the effective mass (shown e.g.\ in Figure~\ref{fig:M(t)}) in (\ref{EQN589}) to generate $\hat{M}_{Q,\hat{T}}$ values for several topological charges $Q = 0,1,2,3,4$ and temporal extensions $\hat{T} = 2.0/\hat{\mathcal{E}}_2 , 3.0/\hat{\mathcal{E}}_2 , \ldots , 10.0/\hat{\mathcal{E}}_2$. Then we perform a single fit of either the expansion (\ref{EQN709}) from \cite{Brower:2003yx} or our $1/V^3$ version restricted to three parameters (eq.\ (\ref{EQN900})) inserted in (\ref{EQN589}) to these masses $\hat{M}_{Q,\hat{T}}$, to determine $\hat{M}_H(0)$ (the hadron mass at unfixed topology), $\hat{M}_H^{(2)}(0)$ and $\hat{\mathcal{E}}_2 = \hat{\chi}_t$ (the curves in Figure~\ref{FIG340}). Only those masses $\hat{M}_{Q,\hat{T}}$ enter the fit, for which the conditions \textbf{(C1)} (we study both $1 / \hat{\mathcal{E}}_2 T , |Q| / \hat{\mathcal{E}}_2 T \leq 0.5$ and $1 / \hat{\mathcal{E}}_2 T , |Q| / \hat{\mathcal{E}}_2 T \leq 0.3$) and \textbf{(C2)} from section~\ref{SEC867} are fulfilled. Both expansions give rather accurate results for $\hat{M}_H(0)$ (cf.\ Table~\ref{TAB002}, top, column ``fitting to $\hat{M}_{Q,\hat{T}}$''; the relative errors are below $0.1 \%$) and reasonable results for $\hat{\chi}_t$ (cf.\ Table~\ref{TAB002}, bottom, column ``fitting to $\hat{M}_{Q,\hat{T}}$''; relative errors of a few percent). Note that the relative errors for both $\hat{M}_H(0)$ and $\hat{\chi}_t$ are smaller, when using the $1/V^3$ version restricted to three parameters (\ref{EQN900}).

\begin{table}[htb]
\begin{center}
$\hat{M}_H(0)$ results from fixed topology computations (exact result: $M_{H} = 0.40714$) \vspace{0.2cm}

\begin{tabular}{|c|c|c|c|c|c|}
\hline 
 & & \multicolumn{2}{c|}{\phantom{xxxx}fitting to $\hat{M}_{Q,\hat{T}}$\phantom{xxxx}} & \multicolumn{2}{c|}{\phantom{xx}fitting to correlators\phantom{xx}}\tabularnewline
\hline 
\hline 
 & expansion & $\hat{M}_H(0)$ result & rel.\ error & $\hat{M}_H(0)$ result & rel.\ error \tabularnewline
\hline 
\multirow{2}{*}{$\frac{1}{\chi_{t}V} , \frac{|Q|}{\chi_{t}V} \leq 0.5$} & hep-lat/0302005 & $0.40733$ & $0.047\%$ & $0.40702$ & $0.029\%$\tabularnewline
\cline{2-6} 
 & (\ref{EQN900}) & $0.40708$ & $0.014\%$ & $0.40706$ & $0.019\%$\tabularnewline
\hline 
\multirow{2}{*}{$\frac{1}{\chi_{t}V} , \frac{|Q|}{\chi_{t}V} \leq 0.3$} & hep-lat/0302005 & $0.40739$ & $0.062\%$ & $0.40732$ & $0.044\%$\tabularnewline
\cline{2-6} 
 & (\ref{EQN900}) & $0.40695$ & $0.046\%$ & $0.40713$ & $0.002\%$\tabularnewline
\hline 
\end{tabular}

\vspace{0.4cm}

$\hat{\chi}_t$ results from fixed topology computations (exact result: $\hat{\chi}_t = 0.00645$) \vspace{0.2cm}

\begin{tabular}{|c|c|c|c|c|c|}
\hline 
 & & \multicolumn{2}{c|}{\phantom{xxxx}fitting to $\hat{M}_{Q,\hat{T}}$\phantom{xxxx}} & \multicolumn{2}{c|}{\phantom{xx}fitting to correlators\phantom{xx}}\tabularnewline
\hline 
\hline 
 & expansion & $\hat{\chi}_t$ result & rel.\ error & $\hat{\chi}_t$ result & rel.\ error \tabularnewline
\hline 
\multirow{2}{*}{$\frac{1}{\chi_{t}V} , \frac{|Q|}{\chi_{t}V} \leq 0.5$} & hep-lat/0302005 & $0.00586$ & $9.1\%$ & $0.00629$ & $2.5\%$\tabularnewline
\cline{2-6} 
 & (\ref{EQN900}) & $0.00631$ & $2.2\%$ & $0.00633$ & $1.9\%$\tabularnewline
\hline 
\multirow{2}{*}{$\frac{1}{\chi_{t}V} , \frac{|Q|}{\chi_{t}V} \leq 0.3$} & hep-lat/0302005 & $0.00590$ & $8.5\%$ & $0.00627$ & $2.8\%$\tabularnewline
\cline{2-6} 
 & (\ref{EQN900}) & $0.00592$ & $8.2\%$ & $0.00630$ & $2.3\%$\tabularnewline
\hline 
\end{tabular}

\caption{\label{TAB002}collection and comparison of results for $\hat{M}_H(0)$ and $\hat{\chi}_t$ from fixed topology computations; ``rel.\ error'' denotes the relative difference to the exact result, i.e.\ the systematic error associated with the determination of $\hat{M}_H(0)$ and $\hat{\chi}_t$ from two-point correlation functions at fixed topology.}

\end{center}
\end{table}

The drawback of this method is that only fixed topology results at a single $t$ value, $t = t_M$, enter the final result for the hadron mass at unfixed topology. To exploit the input data and also the derived $1/V$ expansions for the two-point correlation functions at fixed topology more fully, we propose another method:
\begin{enumerate}
\item Perform simulations at fixed topology for different topological charges $Q$ and spacetime volumes $V$. Determine $C_{Q,V}(t)$ for each simulation.

\item Determine the physical hadron mass $M_H(0)$ by performing a single $\chi^2$ minimizing fit of the preferred $1/V$ expansions of $C_{Q,V}(t)$ (in this work we have discussed nine different versions, (\ref{EQN673}) (\ref{EQN674}), (\ref{EQN708}) and (\ref{EQN709}), (\ref{EQN673V1}), (\ref{EQN673V2}), (\ref{EQN674V1}), (\ref{EQN674V2}), (\ref{EQN900}), (\ref{EQN900_})) with respect to its parameters (cf.\ section~\ref{SEC184} for a detailed summary of available expansions and their parameters) to the two-point correlation functions obtained in step~1. This input from step~1 is limited to those $Q$, $V$ and $t$ values, for which the conditions \textbf{(C1)}, \textbf{(C2)} and \textbf{(C4)} from section~\ref{SEC867} are fulfilled.
\end{enumerate}
Note that this method can also be applied when using correlation matrices at fixed topology. Then corresponding expansions, e.g.\ (\ref{EQN445}) to (\ref{EQN446}), have to fitted simultaneously to all elements of the correlation matrix.

We apply this strategy to the quantum mechanical example using the same $Q = 0,1,2,3,4$ and $\hat{T} = 2.0/\hat{\mathcal{E}}_2 , 3.0/\hat{\mathcal{E}}_2 , \ldots , 10.0/\hat{\mathcal{E}}_2$ values as before. $\hat{t}$ is limited to $12\leqslant \hat{t} \leqslant26$ and sampled equidistantly. Since our input data is exact\footnote{Note that in QCD the exact correlator $C_{Q,V}(t)$ at fixed topological charge $Q$ and spacetime volume $V$ will be provided by lattice simulations, i.e.\ has statistical errors.}, i.e.\ has no statistical errors, the $\chi^2$ minimizing fit becomes and ordinary least squares fit. Again we compare the $1/V$ expansion from \cite{Brower:2003yx} (eqs.\ (\ref{EQN708}) and (\ref{EQN709})) and our $1/V^3$ version restricted to three parameters (\ref{EQN900}). As before, we find rather accurate results for $\hat{M}_H(0)$ and $\hat{\chi}_t$ (cf.\ Table~\ref{TAB002}, columns ``fitting to correlators''). Note that the relative errors for both $\hat{M}_H(0)$ and $\hat{\chi}_t$ are smaller, when using the $1/V^3$ version restricted to three parameters (\ref{EQN900}). The relative errors are also smaller compared to the previously discussed method of ``fitting to $\hat{M}_{Q,\hat{T}}$''.


\section{Conclusions and outlook}

In this work we have extended a calculation of the $Q$, $V$ and $t$ dependence of two-point correlation functions at fixed topology from \cite{Brower:2003yx}. While in \cite{Brower:2003yx} the expansion included all terms of $\mathcal{O}(1/\chi_t V)$ and some of $\mathcal{O}(1/(\chi_t V)^2)$, we have derived the complete result up to $\mathcal{O}(1/(\chi_t V)^3)$. Since $\chi_t V \ltapprox 10$ in many ensembles of typical nowadays lattice QCD simulations (cf.\ e.g.\ \cite{Aoki:2007pw,Chiu:2011dz,Cichy:2013rra,Brower:2014bqa}), fixed topology corrections of order $1/(\chi_t V)^2$ or even $1/(\chi_t V)^3$ might be sizable, in particular for topological charge $Q \geq 2$, as e.g.\ demonstrated in Figure~\ref{FIG199}. We have also discussed parity mixing in detail, which appears at fixed topology already at $\mathcal{O}(1/\chi_t V)$. In particular we have derived corresponding expansions of correlation functions between $P = -$ and $P = +$ operators as well as contributions of opposite parity hadrons to correlation functions between operators of identical parity.

We have applied, discussed and checked our results in the context of a simple model, a quantum mechanical particle on a circle, both in the free case and for a square well potential. We have studied and compared various orders and versions of the $1/V$ expansion of $C_{Q,V}(t)$ differing in accuracy and in the number of parameters. We also discussed and demonstrated, how to extract a mass at unfixed topology from computations of two-point correlation functions at fixed topology. In practice, e.g.\ in QCD, where computed two-point correlation functions have limited accuracy, due to statistical errors, one probably needs a $1/V$ expansion of $C_{Q,V}(t)$ with a rather small number of parameters to be able to perform a stable fit. We recommend to use the $1/V$ expansion (\ref{EQN900}), which seems to be a good compromise:
\begin{itemize}
\item It contains certain $1/V^2$ and $1/V^3$ terms and, therefore, seems to be more accurate than the expansion from \cite{Brower:2003yx} (cf.\ Figure~\ref{FIG199}).

\item At the same time the number of fit parameters is quite small ($\mathcal{E}_2$, $M_H(0)$ and $M_H^{(2)}(0)$), the same as for the expansion from \cite{Brower:2003yx}.
\end{itemize}

Currently we are applying the equations and methods derived and discussed in this work to simple quantum field theories, e.g.\ the Schwinger model and pure Yang-Mills theory (cf.\ also \cite{Bietenholz:2011ey,Bietenholz:2012sh,Czaban:2013haa,Bautista:2014tba,Czaban:2014gva} for existing work in this direction). The final goal is, of course, to develop and establish methods to reliably extract hadron masses from QCD simulations at fixed topology.


\appendix

\section{Technical aspects of a quantum mechanical particle on a circle in a square well}


\subsection{\label{SEC634}Wave functions}

After replacing $p_\varphi \rightarrow -i \partial_\varphi$, Schr\"odinger's equation is
\begin{eqnarray}
\bigg(\frac{1}{2 I} \bigg(-i \partial_\varphi + \frac{\theta}{2 \pi}\bigg)^2 + U(\varphi)\bigg) \psi_n(\varphi;\theta) = E_n(\theta) \psi_n(\varphi;\theta) .
\end{eqnarray}
The wave function with energy $E_n(\theta)$ in ``region 1'', $-\rho/2 < \varphi < +\rho/2$, where $U(\varphi) = 0$, is
\begin{equation}
\psi_n^{(1)}(\varphi;\theta) = \Big(A_n(\theta) e^{+i p \varphi} + B_n(\theta) e^{-i p \varphi}\Big) e^{-i (\theta / 2 \pi)\varphi} \quad , \quad p = \sqrt{2 E_n I} ,
\end{equation}
in ``region 2'', $+\rho/2 < \varphi < 2 \pi - \rho/2$, where $U(\varphi) = U_0$,
\begin{equation}
\psi_n^{(2)}(\varphi;\theta) = \Big(C_n(\theta) e^{+i q \varphi} + D_n(\theta) e^{-i q \varphi}\Big) e^{-i (\theta / 2 \pi)\varphi} \quad , \quad q = \sqrt{2 (E_n - U_0) I} .
\end{equation}
The coefficients $A_n(\theta)$, $B_n(\theta)$, $C_n(\theta)$ and $D_n(\theta)$ have to be chosen such that both the wave function and its derivative are continuous, i.e.\ that
\begin{equation}
\begin{aligned}
 & \psi_n^{(1)}(+\rho/2;\theta) = \psi_n^{(2)}(+\rho/2;\theta) \quad , \quad \psi_n^{(2)}(2\pi-\rho/2;\theta) = \psi_n^{(1)}(-\rho/2;\theta) \\
 & \psi'{}_n^{(1)}(+\rho/2;\theta) = \psi'{}_n^{(2)}(+\rho/2;\theta) \quad , \quad \psi'{}_n^{(2)}(2\pi-\rho/2;\theta) = \psi'{}_n^{(1)}(-\rho/2;\theta) \\
\end{aligned}
\end{equation}
are fulfilled, which is only possible for specific discrete values of $E_n(\theta)$. Note that, even after properly normalizing the wave function $\psi_n(\varphi;\theta)$, its coefficients $A_n(\theta)$, $B_n(\theta)$, $C_n(\theta)$ and $D_n(\theta)$ are only unique up to a phase.


\subsection{\label{SEC635}Probability density to find a particle}

The probability density to find a particle with wave function $\psi_n(\varphi;\theta)$ is $P_n(\varphi;\theta) \equiv |\psi_n(\varphi;\theta)|^2$. In the following it will be shown that $P_n(+\varphi;\theta) = P_n(-\varphi;\theta)$.

First note that $(\psi_n(\varphi;\theta))^\ast$ and $\psi_n(-\varphi;\theta)$ fulfill the same Schr\"odinger equation, which implies
\begin{equation}
\label{EQN796} (\psi_n(\varphi;\theta))^\ast = \eta \psi_n(-\varphi;\theta) ,
\end{equation}
where $\eta$ is a non-unique phase.

Now consider region~1, where
\begin{equation}
(\psi_n(\varphi;\theta))^\ast = \Big((A_n(\theta))^\ast e^{-i p \varphi} + (B_n(\theta))^\ast e^{+i p \varphi}\Big) e^{+i (\theta / 2 \pi)\varphi}
\end{equation}
and
\begin{equation}
\psi_n(-\varphi;\theta) = \Big(A_n(\theta) e^{-i p \varphi} + B_n(\theta) e^{+i p \varphi}\Big) e^{+i (\theta / 2 \pi)\varphi} .
\end{equation}
Inserting these expressions in (\ref{EQN796}) yields
\begin{equation}
(A_n(\theta))^\ast = \eta A_n(\theta) \quad , \quad (B_n(\theta))^\ast = \eta B_n(\theta)
\end{equation}
and, consequently,
\begin{equation}
A_n(\theta) (B_n(\theta))^\ast = (A_n(\theta))^\ast B_n(\theta) .
\end{equation}
With this relation it is easy to show that the probability density is an even function,
\begin{equation}
\begin{aligned}
 & P_n(+\varphi;\theta) = (\psi_n(+\varphi;\theta))^\ast \psi_n(+\varphi;\theta) = \\
 & \hspace{0.7cm} = |A_n(\theta)|^2 + |B_n(\theta)|^2 + \underbrace{A_n(\theta) (B_n(\theta))^\ast}_{= (A_n(\theta))^\ast B_n(\theta)} e^{+2 i p \varphi}  + \underbrace{(A_n(\theta))^\ast B_n(\theta)}_{= A_n(\theta) (B_n(\theta))^\ast} e^{-2 i p \varphi} = \\
 & \hspace{0.7cm} = (\psi_n(-\varphi;\theta)^\ast \psi_n(-\varphi;\theta) = P_n(-\varphi;\theta) .
\end{aligned}
\end{equation}

Using similar arguments one can show that also in region~2 $P_n(+\varphi;\theta)$ is an even function.

An important consequence is
\begin{equation}
\langle 0 ; \theta | \sin(\varphi) | 0 ; \theta \rangle = \int_0^{2 \pi} d\varphi \, (\psi_n(\varphi;\theta))^\ast \sin(\varphi) \psi_n(\varphi;\theta) = \int_0^{2 \pi} d\varphi \, \underbrace{P_n(\varphi;\theta)}_{\textrm{even}} \underbrace{\sin(\varphi)}_{\textrm{odd}} = 0 ,
\end{equation}
which has been used in section~\ref{SEC694}.


\section*{Acknowledgments}

We thank Wolfgang Bietenholz, Krzysztof Cichy, Christopher Czaban, Dennis Dietrich, Gregorio Herdoiza, Karl Jansen and Andreas Wipf for discussions.

We acknowledge support by the Emmy Noether Programme of the DFG (German Research Foundation), grant WA 3000/1-1.

This work was supported in part by the Helmholtz International Center for FAIR within the framework of the LOEWE program launched by the State of Hesse.




\begin{thebibliography}{99}

\bibitem{Kennedy:2006ax}
  A.~D.~Kennedy,
  hep-lat/0607038.

\bibitem{Luscher:2011kk}
  M.~L\"uscher and S.~Schaefer,
  JHEP {\bf 1107}, 036 (2011)
  [arXiv:1105.4749 [hep-lat]].

\bibitem{Schaefer:2012tq}
  S.~Schaefer,
  PoS LATTICE {\bf 2012}, 001 (2012)
  [arXiv:1211.5069 [hep-lat]].

\bibitem{Aoki:2008tq} 
  S.~Aoki {\it et al.} [JLQCD Collaboration],
  Phys.\ Rev.\ D {\bf 78}, 014508 (2008)
  [arXiv:0803.3197 [hep-lat]].

\bibitem{Aoki:2012pma}
  S.~Aoki, T.~-W.~Chiu, G.~Cossu, X.~Feng, H.~Fukaya, S.~Hashimoto, T.~-H.~Hsieh and T.~Kaneko {\it et al.},
  PTEP {\bf 2012}, 01A106 (2012).
  
\bibitem{Galletly:2006hq} 
  D.~Galletly, M.~Gurtler, R.~Horsley, H.~Perlt, P.~E.~L.~Rakow, G.~Schierholz, A.~Schiller and T.~Streuer,
  Phys.\ Rev.\ D {\bf 75}, 073015 (2007)
  [hep-lat/0607024].

\bibitem{Fukaya:2005cw}
  H.~Fukaya, S.~Hashimoto, T.~Hirohashi, K.~Ogawa and T.~Onogi,
  Phys.\ Rev.\ D {\bf 73}, 014503 (2006)
  [hep-lat/0510116].

\bibitem{Bietenholz:2005rd}
  W.~Bietenholz, K.~Jansen, K.~-I.~Nagai, S.~Necco, L.~Scorzato and S.~Shcheredin,
  JHEP {\bf 0603}, 017 (2006)
  [hep-lat/0511016].

\bibitem{Bruckmann:2009cv} 
  F.~Bruckmann, F.~Gruber, K.~Jansen, M.~Marinkovic, C.~Urbach and M.~Wagner,
  Eur.\ Phys.\ J.\ A {\bf 43}, 303 (2010)
  [arXiv:0905.2849 [hep-lat]].

\bibitem{Cichy:2010ta}
  K.~Cichy, G.~Herdoiza and K.~Jansen,
  Nucl.\ Phys.\ B {\bf 847}, 179 (2011)
  [arXiv:1012.4412 [hep-lat]].

\bibitem{Cichy:2012vg}
  K.~Cichy, V.~Drach, E.~Garcia-Ramos, G.~Herdoiza and K.~Jansen,
  Nucl.\ Phys.\ B {\bf 869}, 131 (2013)
  [arXiv:1211.1605 [hep-lat]].

\bibitem{Brower:2003yx} 
  R.~Brower, S.~Chandrasekharan, J.~W.~Negele and U.~J.~Wiese,
  Phys.\ Lett.\ B {\bf 560}, 64 (2003)
  [hep-lat/0302005].

\bibitem{Aoki:2007ka}
  S.~Aoki, H.~Fukaya, S.~Hashimoto and T.~Onogi,
  Phys.\ Rev.\ D {\bf 76}, 054508 (2007)
  [arXiv:0707.0396 [hep-lat]].

\bibitem{Aoki:2007pw} 
  S.~Aoki {\it et al.} [JLQCD and TWQCD Collaborations],
  Phys.\ Lett.\ B {\bf 665}, 294 (2008)
  [arXiv:0710.1130 [hep-lat]].

\bibitem{Chiu:2011dz} 
  T.~W.~Chiu, T.~H.~Hsieh and Y.~Y.~Mao,
  Phys.\ Lett.\ B {\bf 702}, 131 (2011)
  [arXiv:1105.4414 [hep-lat]].

\bibitem{Cichy:2013rra} 
  K.~Cichy {\it et al.} [ETM Collaboration],
  JHEP {\bf 1402}, 119 (2014)
  [arXiv:1312.5161 [hep-lat]].

\bibitem{Brower:2014bqa} 
  R.~C.~Brower {\it et al.} [LSD Collaboration],
  arXiv:1403.2761 [hep-lat].



\bibitem{Bietenholz:2011ey}
  W.~Bietenholz, I.~Hip, S.~Shcheredin and J.~Volkholz,
  Eur.\ Phys.\ J.\ C {\bf 72}, 1938 (2012)
  [arXiv:1109.2649 [hep-lat]].

\bibitem{Bietenholz:2012sh}
  W.~Bietenholz and I.~Hip,
  J.\ Phys.\ Conf.\ Ser.\  {\bf 378}, 012041 (2012)
  [arXiv:1201.6335 [hep-lat]].

\bibitem{Czaban:2013haa}
  C.~Czaban and M.~Wagner,
  arXiv:1310.5258 [hep-lat].

\bibitem{Bautista:2014tba}
  I.~Bautista, W.~Bietenholz, U.~Gerber, C.~P.~Hofmann, Héc.~Mejía-Díaz and L.~Prado,
  arXiv:1402.2668 [hep-lat].

\bibitem{Dromard:2013wja}
  A.~Dromard and M.~Wagner,
  arXiv:1309.2483 [hep-lat].

\bibitem{Czaban:2014gva} 
  C.~Czaban, A.~Dromard and M.~Wagner,
  arXiv:1404.3597 [hep-lat].

\bibitem{Coleman:1978ae}
  S.~R.~Coleman,
  Subnucl.\ Ser.\ {\bf 15}, 805 (1979).

\bibitem{Bonati:2013tt}
  C.~Bonati, M.~D'Elia, H.~Panagopoulos and E.~Vicari,
  Phys.\ Rev.\ Lett.\ {\bf 110}, 252003 (2013)
  [arXiv:1301.7640 [hep-lat]].

\bibitem{Weber:2013eba} 
  J.~Weber, S.~Diehl, T.~Kuske and M.~Wagner,
  arXiv:1310.1760 [hep-lat].

\bibitem{Jansen:2008si} 
  K.~Jansen {\it et al.} [ETM Collaboration],
  JHEP {\bf 0812}, 058 (2008)
  [arXiv:0810.1843 [hep-lat]].

\bibitem{Michael:2010aa} 
  C.~Michael, A.~Shindler and M.~Wagner [ETM Collaboration],
  JHEP {\bf 1008}, 009 (2010)
  [arXiv:1004.4235 [hep-lat]].

\bibitem{Baron:2010th} 
  R.~Baron {\it et al.} [ETM Collaboration],
  Comput.\ Phys.\ Commun.\  {\bf 182}, 299 (2011)
  [arXiv:1005.2042 [hep-lat]].

\bibitem{Wagner:2011fs} 
  M.~Wagner and C.~Wiese [ETM Collaboration],
  JHEP {\bf 1107}, 016 (2011)
  [arXiv:1104.4921 [hep-lat]].

\bibitem{Kalinowski:2012re} 
  M.~Kalinowski and M.~Wagner [ETM Collaboration],
  PoS ConfinementX {\bf }, 303 (2012)
  [arXiv:1212.0403 [hep-lat]].

\bibitem{Kalinowski:2013wsa} 
  M.~Kalinowski and M.~Wagner [ETM Collaboration],
  Acta Phys.\ Polon.\ Supp.\  {\bf 6}, no. 3, 991 (2013)
  [arXiv:1304.7974 [hep-lat]].

\bibitem{Wagner:2013laa} 
  M.~Kalinowski and M.~Wagner [ETM Collaboration],
  PoS LATTICE {\bf 2013}, 241 (2013)
  [arXiv:1310.5513 [hep-lat]].

\end{thebibliography}
\end{document}